# *In situ* and real-time ultrafast spectroscopy of photoinduced reactions in perovskite nanomaterials

Gi Rim Han[1], Mai Ngoc An[1,a], Hyunmin Jang[1,2], Noh Soo Han[1,b], JunWoo Kim[1,c] , Kwang Seob Jeong[1,2], Tai Hyun Yoon[1,3,*], Minhaeng Cho[1,2,*]

[1]Center for Molecular Spectroscopy and Dynamics, Institute for Basic Science (IBS); Seoul, 02841, Republic of Korea.
[2]Department of Chemistry, Korea University; Seoul, 02841, Republic of Korea.

[3]Department of Physics, Korea University, Seoul, 02841, Republic of Korea.

*Corresponding authors. Email: mcho@korea.ac.kr (MC) or thyoon@korea.ac.kr (THY)

[a]Present address: Quantum Science Ltd. TechSpace One, Sci-Tech Daresbury, Keckwick Lane, Warrington, WA4 4AB, United Kingdom.

[b]Present Address: Korea Research Institute of Standards and Science, Daejeon, 34113, Republic of Korea.

[c]Present Address: Department of Chemistry, Chungju, Chungcheongbuk-do, 28864, Republic of Korea.

**Employing two synchronized mode-locked femtosecond lasers and interferometric detection of the pump-probe spectra—referred to as asynchronous and interferometric transient absorption (AI-TA)—we have developed a method for broad dynamic range and rapid data acquisition. Using AI-TA, we examined photochemical changes during femtosecond pump-probe experiments on all-inorganic cesium lead halide nanomaterials, including perovskite nanocrystals (PeNCs) and nanoplatelets (PeNPLs). The laser pulse train facilitates photoreactions while allowing real-time observation of charge carrier dynamics. In PeNCs undergoing halide anion photo-substitution, transient absorption spectra showed increasing bandgap energy and faster relaxation dynamics as the Cl/Br ratio increased. For colloidal PeNPLs, continuous observation revealed both spectral and kinetic changes during the light-induced coalescence of nanoplatelets, by analyzing temporal segments. This integrated technique not only deepens understanding of exciton dynamics and environmental influences in perovskite nanomaterials but also establishes AI-TA as a transformative tool for real-time observation of photochemical dynamics.**

## Main

In spectroscopy, researchers frequently encounter a dilemma when the tool they employ for observation – light – also triggers alterations in their target materials and molecular systems. This problem is particularly pronounced for ultrafast spectroscopy methods. Here, prolonged exposure to high-energy laser light is imperative to capture subtle and weak nonlinear responses. The situation underscores the challenge: the acquisition of data under such circumstances becomes demanding and, at times, seemingly impossible. Therefore, the reduction of data acquisition (DAQ) time is essential when the observed materials deviate from established photostability standards.

The efficacy of asynchronous optical sampling in curtailing DAQ time is well-established, as it eliminates the need for linear stages to generate time delays between pump and probe pulses[1,2]. Although initially proposed and demonstrated nearly half a century ago, our group has been in pursuit of advancing this technique across various bandwidth ranges and applied to a wide array of biological and material systems[3–6]. However, the broader research community in ultrafast spectroscopy remains incompletely cognizant of its utility for rapid and stable transient absorption and multidimensional spectroscopic measurements. In this work, we introduce and apply our novel spectroscopic approach, asynchronous and interferometric transient absorption (AI-TA), to investigate the photochemical reactions of perovskite nanocrystals (PeNCs) and nanoplatelets (PeNPLs).

Perovskite nanomaterials have attracted considerable attention for their applications in light-emitting diodes[7] and as single-photon sources in quantum technologies[8]. Despite their susceptibility to trap and defect formations due to a high surface-to-volume ratio in colloidal dispersions[9,10], these materials exhibit advantageous properties, including high defect tolerance[11], tunable bandgap levels[12], and a simple, cost-effective synthesis process[13]. Additionally, they possess notable physical attributes such as high quantum yields[14,15] and enhanced stability[16], making them promising candidates for photovoltaic and optoelectronic devices.

Given their potential applications in solar cells and light-emitting devices, the interaction of perovskites with light has become a subject of significant interest among researchers and engineers. Beyond time-resolved studies of carrier and quasiparticle dynamics, perovskite nanomaterials display a range of light-induced behaviors, including degradation, which has emerged as a key focus in recent research. Here, to demonstrate the rapid DAQ capabilities of AI-TA, we simultaneously investigate both ultrafast carrier dynamics via femtosecond pump-probe measurements and light-induced transformations in perovskite nanomaterials, where two distinct types of photoreactions are considered.

The first study examines photoinduced halide substitution of PeNCs. A key feature of PeNCs is their possibility of post-synthesis compositional modification through the use of appropriate salts containing halide anions, as they can freely diffuse into and out of the rigid cation lattice[17]. The halide composition significantly influences the bandgap energy, allowing the emission wavelength to be tuned across the visible spectrum. For instance, iodide-based PeNCs exhibit red to near-infrared emission, while bromide and chloride shift the bandgap toward blue and near-ultraviolet regions[12].

Recent studies have reported halide anion substitution in perovskites through a photoinduced process[18,19]. This phenomenon can take place when PeNCs are dissolved in halide-containing solvents (e.g., chloroform and dibromomethane) and concomitantly exposed to light. The proposed mechanism involves the dissociation of solvent molecules upon interaction with photoexcited PeNCs, facilitated by interfacial electron transfer[18,20]. The high reduction potential of these solvents relative to PeNCs enables photoinduced electron transfer, extracting halide anions from the solvents. These anions can then exchange and diffuse within the PeNC lattice. The photoinduced anion exchange process allows remote control over the extent and composition of substitution[19] while preserving the size and morphology of the nanoparticles[21]. Additionally, it has been shown to mitigate vacancy defects, enhance crystallinity[22], and facilitate cation exchanges as well[23], further broadening the potential applications of PeNCs.

The second system under investigation involves photon-driven transformations of PeNPLs. PeNPLs represent a distinct class of nanomaterials with unique morphology and physical properties compared to PeNCs. Their defining characteristic arises from confinement along a single-dimensional axis, with thinner platelets exhibiting higher bandgap energy[24]. However, their colloidal stability in solution is limited, and PeNPLs are known to aggregate into higher-order structures, including bulk-like status, when exposed to heat[25,26], light[27–30], polar solvents[31], or prolonged ambient conditions[32]. This phenomenon, particularly under light exposure, is referred to as photon-driven transformation.

Such self-assembly is not exclusive to PeNPLs and can also occur in other colloidal forms, including PeNCs and perovskite nanorods[33]. However, PeNPLs are especially prone to aggregation into higher-dimensional structures. This process often begins with pre-assembly into self-assembled superlattices, where discrete nanoplatelets stack with ligands spacing the ionic crystal layers[33]. These intermediate assemblies may aggregate into larger structures, sometimes forming mosaic-like superstructures[25,26]. In this second study, the AI-TA technique is applied to track real-time changes occurring in PeNPLs during the pump-probe measurement, without the deliberate introduction of reactants as in the previous case. Time-resolved *in situ* observations reveal coalescence-induced changes in the transient absorption spectra of PeNPLs, demonstrating the technique’s capability for monitoring light-induced dynamic transformations of perovskite nanomaterials.

In this Communication, we begin by providing an overview of our approach, AI-TA, emphasizing its capability to enable the distinct observations presented in this study. We then analyze the transient dynamics and spectral features of PeNCs dispersed in toluene solvent under non-reactive conditions. Subsequently, we explore the *in situ* photoinduced substitution process, examining changes in the bromide-to-chloride ratio and their effects on the ultrafast transient absorption spectra of charge carriers. For the next target, we investigate PeNPL colloidal dispersions using AI-TA, tracking spectral and temporal changes as the nanoplatelets aggregate into higher-order structures. In addition to presenting novel photophysical insights from these two experiments, we demonstrate that AI-TA is a powerful technique for rapid DAQ, enabling *in situ* ultrafast spectroscopy to monitor processes occurring over extended timescales.

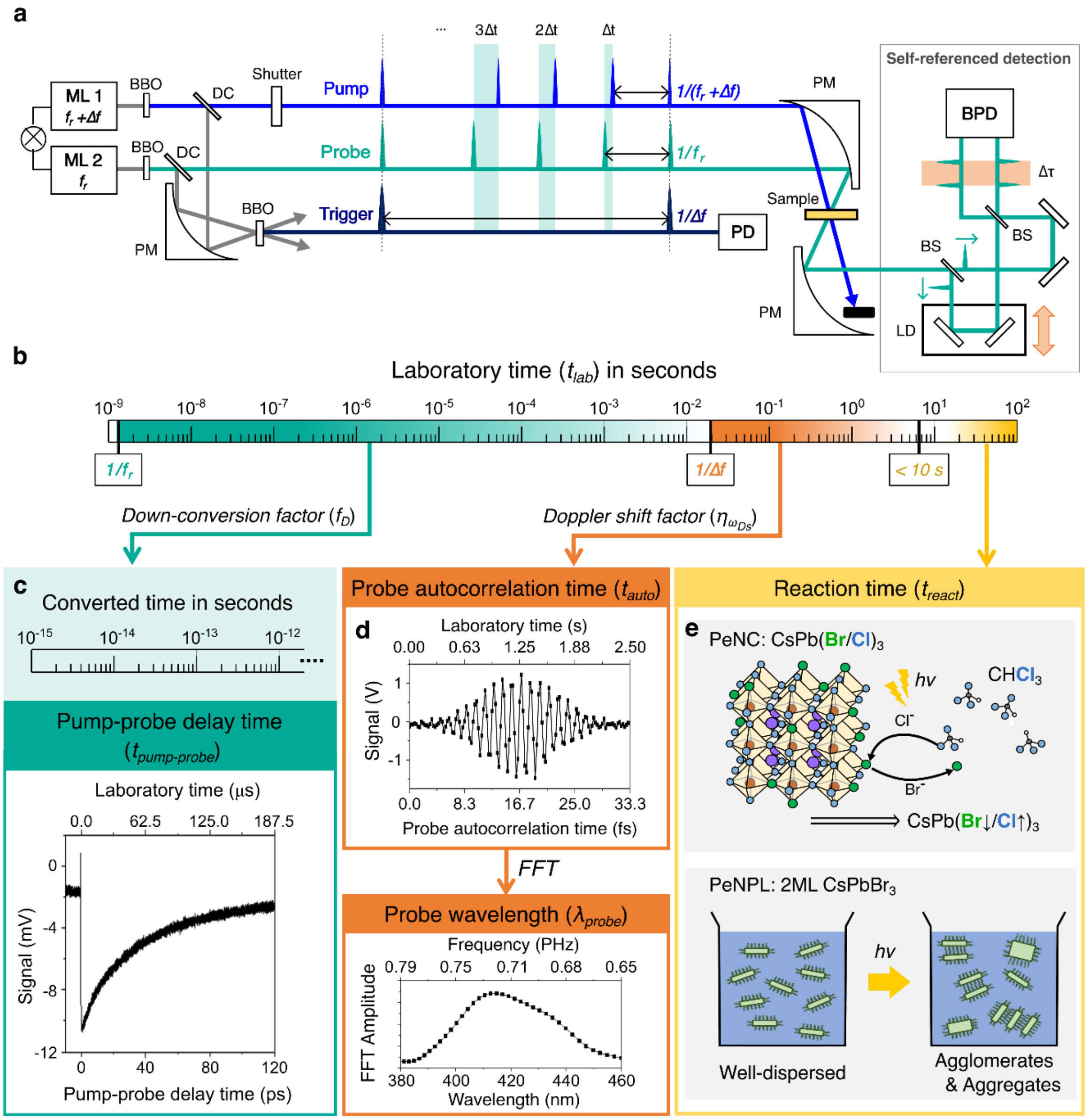

**Fig. 1. Schematics of AI-TA instrumentation and parameters.** (**a**) Experimental setup of the AI-TA. The spectra of ML1 and ML2 are shown in Fig. S1. ML: mode-locked laser, DC: dichroic mirror, BBO: barium-borate crystal. PD: photodiode, DC: dichroic mirror, BS: beam-splitter, PM: parabolic mirror, BPD: balanced photodiode, LD: linear delay stage. The self-referenced interferogram is acquired through a Mach-Zehnder configuration and detected using a balanced photodetector. (**b**) Definition of the time variables in relation to the laboratory time, $t_{lab}$. By using time-division multiplexing, the laboratory time can be processed into pump-probe delay time ($t_{pump-probe}$), probe autocorrelation time ($t_{auto}$), and reaction time ($t_{react}$). (**c**) An example of a transient signal depending on the $t_{pump\text{-}probe}$, which is converted from the laboratory time using the down-conversion factor, $f_D$. Here, when $f_r$=80.0 MHz and $\Delta f$=51.2 Hz, 187.5 μs of laboratory time is required to complete a single scan of pump-probe delay from 0 to 120 ps. (**d**) The interferogram acquired by collecting signals sampled at the rate of $\Delta f$=51.2 Hz, while the linear delay stage is moving at a constant velocity of 2 μm/s. The laboratory time in this case can be converted to the $t_{auto}$ using the Doppler shift factor $\eta_{\omega DS}$. Fast Fourier transform of the data yields the probe spectrum with respect to the wavelength, $\lambda_{probe}$. (**e**) Experimental schematic of the ultrafast pump-probe investigation of photoinduced reactions of perovskite nanomaterials. Since the pump and the probe light work as reaction stimulants that induce photo-substitution or photon-driven transformation, the data acquiring measurement time can be considered as $t_{react}$.

**Asynchronous and interferometric transient absorption (AI-TA) spectroscopy**

Numerous attempts have been made to reduce the DAQ time associated with transient absorption, one of the most widely used third-order nonlinear spectroscopy techniques. Reflecting on past initiatives, such as the adoption of array detectors and computational data reconstruction, we can identify two key rate-determining factors[1,34,35]: first, the generation of pump-probe time delays using linear stages, and second, the execution of data averaging to reduce noise levels.

AI-TA overcomes the limitations imposed by conventional techniques by eliminating the need for a linear stage to generate pump-probe delays. Instead, it employs a pair of repetition rate-stabilized mode-locked lasers. When the two lasers have slightly different repetition rates, they generate time steps proportional to the difference in repetition rates (Fig. 1a), referred to as the detuning frequency ($\Delta f$). This method ensures consistent spatial overlap of the pump and probe beams, regardless of the duration of the time delays. The equation below shows how to calculate each time step ($\Delta t$) generated by a pair of pump and probe pulses with repetition rates, $f_r$ and $f_r + \Delta f$, respectively:

$$\Delta t = \frac{1}{f_r} - \frac{1}{f_r + \Delta f} = \frac{\Delta f}{f_r(f_r + \Delta f)} \approx \frac{\Delta f}{f_r^2} = \frac{f_D}{f_r}. \qquad (1)$$

The approximation in equation (1) is valid due to the significantly small detuning frequency $\Delta f$ compared to the repetition frequency ($f_r$), i.e., $\Delta f \ll f_r$. An important point to note is that the down-conversion factor $f_D = \Delta f / f_r$ can be precisely controlled by a GPS-disciplined Rb atomic clock, ensuring consistent generation of time steps[36,37]. Additionally, when the separated pump and probe pulses overlap in space and time on another BBO nonlinear crystal for sum-frequency generation, distinct trigger signals are produced at a rate of $1/\Delta f$, marking the initiation of the pump-induced probe response.

In conventional transient absorption techniques, obtaining wavelength-resolved information in the time domain requires a monochromator or spectrograph equipped with a charge-coupled device (CCD). In our approach, we accomplish this by generating a self-referenced interferogram (see Fig. S1 for the pump and probe spectra). The position of the linear stage corresponds to the time delay between split probe pulses, so scanning the stage at a constant velocity produces an autocorrelation signal (Fig. S2). Using frequency domain detection with a single point detector offers advantages in terms of effective noise filtration and signal collection efficiency.

In our setup, the probe intensity is detected by a fast digitizer operating at a sampling rate of $f_r$, synchronized with the repetition rate of the probe laser. The probe signal contains information about both the linear absorbance of the sample and its pump-induced nonlinear responses. Since the raw data consists of an array of voltage signals that are time-tagged by optical trigger signals, time-division multiplexing can be applied to acquire information from segments of different temporal lengths[3,37]. In Fig. 1b, to avoid confusion in the upcoming discussions, we define four temporal variables involved in the interpretation of AI-TA data. The time tag from the digitizer represents the real-world time measured during the experiment, also referred to as laboratory time ($t_{lab}$). Given that our system records data every 12.5 ns, the temporal span over a one-hour experiment encompasses eleven orders of temporal dynamic range. This extensive time span allows the analysis of additional time-dependent variables that influence the probe response if they operate in temporally distinct regimes. Based on this, laboratory time is divided into three segments in this study: pump-probe delay time, probe autocorrelation time, and reaction time.

First and foremost, using the relationship (Eq. 1), we can convert the laboratory time, recorded at an interval of $1/f_r$ to the pump-probe delay time ($t_{pump\text{-}probe}$) by multiplying it with the down-conversion factor $f_D$. In Fig. 1c, the pump-probe response of the PeNC solution is shown as a function of $t_{pump\text{-}probe}$. The starting time, or time zero, of the signal is defined by the optical trigger signal. Then, the pump-induced transmission change is calculated by dividing the voltage signals recorded after the trigger by those recorded before the trigger.

The interferogram generated by scanning the linear stage contains information about the frequency of the electromagnetic field, which can be easily accessed through a fast Fourier transform (FFT). The method requires the data points of the interferogram to be evenly spaced in time. Provided that the linear stage moves at a fixed velocity, this can be achieved by using the trigger rate, $\Delta f$, as the sampling frequency, as shown in Fig. 1d. The time interval between each sample ($\Delta\tau$) can be calculated using the Doppler shift relation $\omega_{Ds} = 2\omega v/c,$ where $\omega$ is the angular frequency of the probe field, $v$ is the stage speed, and $c$ denotes the speed of light. Then, we can deduce that $\Delta\tau = 2v/c\Delta f = \eta_{\omega_{Ds}}/\Delta f$. The Doppler shift factor, $\eta_{\omega_{Ds}} = 2v/c$, is multiplied by $t_{lab}$ to calculate the effective time between split probe pulses, known as the probe autocorrelation time ($t_{auto}$).

From here, we can extract frequency-resolved light intensity via FFT, which can also be expressed in terms of probe wavelength ($\lambda_{probe}$) using the speed of light $c$. Since each data point is annotated with a corresponding $t_{lab}$ in relation to trigger time zero, our approach allows simultaneous measurement of decay lifetimes for each spectral component, provided the linear stage scanning speed is significantly slower than the rate of pump-probe delay generation time $t_{pump\text{-}probe}$. More detailed theoretical descriptions and comprehensive analyses of experimental parameters related to this technique have been presented and demonstrated in our previous works[3,38]. Our implementation, AI-TA, also features a high duty cycle and low energy per pulse, rendering it particularly advantageous for scenarios involving photoreactions and photodamages.

Figure 1e presents two target systems for the ultrafast pump-probe investigation in this study and defines the final time parameter, i.e., reaction time ($t_{react}$). This experimental design is conceived to integrate both synthesis and analysis, with the pump and probe light serving as both reaction stimulants and observation tools simultaneously. Meanwhile, the rapid DAQ procedure of our AI-TA setup enables data collection before significant changes occur in the perovskite nanomaterials. Since the data from each linear stage scan is saved as a discrete file with a timestamp, we can determine the interval between subsequent measurements, which is less than 10 s (Fig. S3). Thus, this measurement time can be considered the reaction time in our analysis. Note that $t_{react}$ has fewer significant digits, as it is not governed by the precision of an atomic clock.

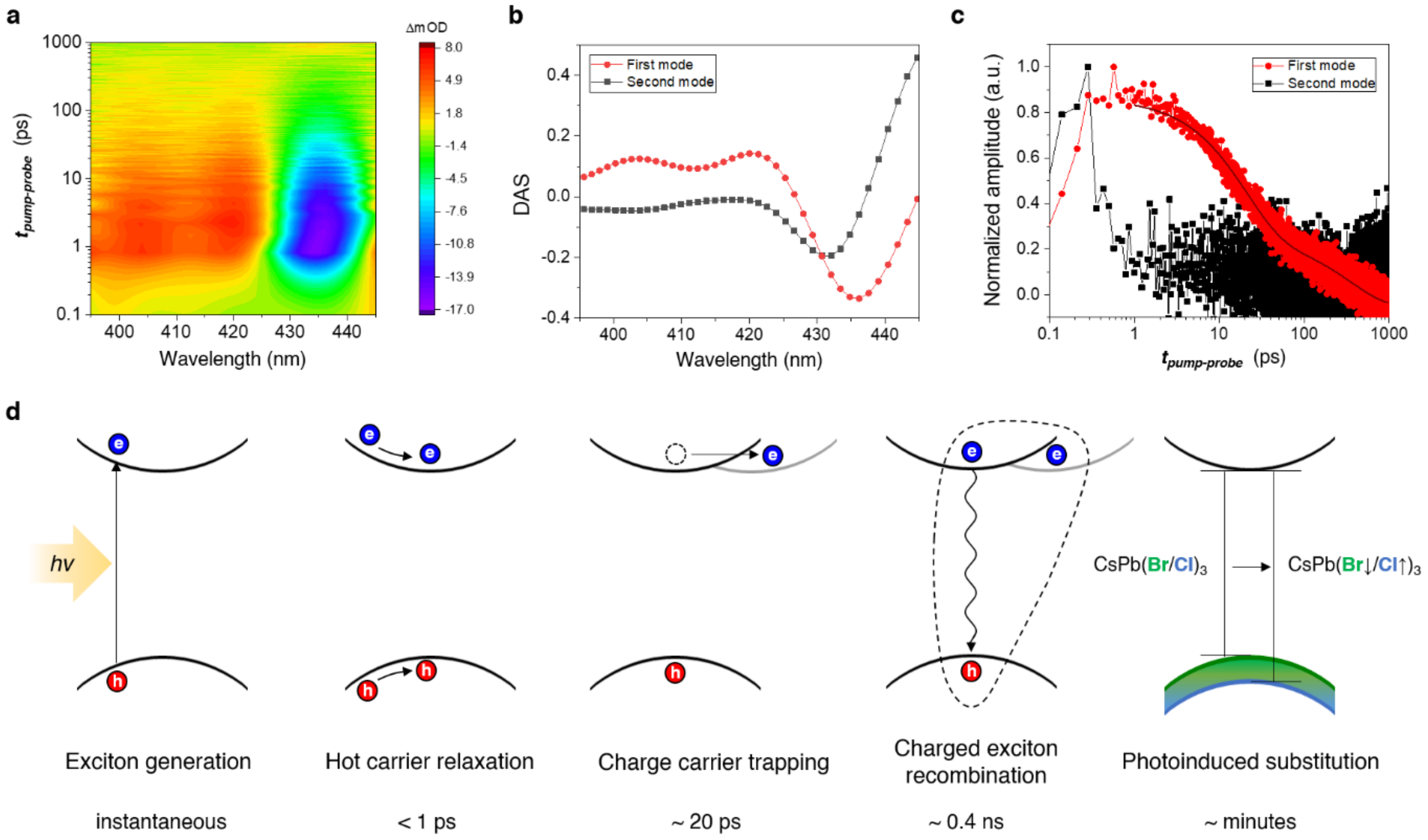

**Fig. 2. Photodynamics and photo-substitution of PeNCs.** (**a**) The transient absorption spectrum of $CsPb(Br/Cl)_3$ PeNCs dissolved in toluene measured up to 1 ns. The spectrum is the average result of 200 scans. (**b**) Decay-associated spectra and (**c**) time-resolved decay profiles of the first two modes from singular value decomposition analysis. The second mode decays quickly within 1 ps, and the first mode persists over a longer period, which has been analyzed with a bi-exponential fit. (**d**) List of various dynamics observable with the AI-TA technique spanning a broad dynamic range from sub-picosecond to minutes. The initial pump pulse instantaneously generates holes and electrons. As the excitation energy exceeds the bandgap energy, any excess energy is quickly dissipated as the hot carriers relax to the band-edge state. This transition is accompanied by bandgap renormalization. Due to the presence of trap states near the band edge, the conduction band is partially depopulated with a short lifetime of approximately 20 ps. The shallow nature of this charge trap allows interaction with other excitons, resulting in a charged exciton recombination with a lifetime of 0.35 ns.

## AI-TA measurement of $CsPb(Br/Cl)_3$ in toluene

Before proceeding to *in situ* analysis, we first analyzed the transient absorption spectrum of the nanocrystals dissolved in toluene to investigate the dynamics involved in the relaxation of PeNCs. Since this solvent does not contain halide agents, halide exchange is avoided. The resulting TA spectrum, shown in Fig. 2a, represents an average of 200 interferogram scans. We applied singular value decomposition (SVD) and selected the first two eigenmodes as the principal representations of the states contributing to the overall transient absorption signal (Fig. S4).

The decay-associated spectrum (DAS) of the first mode exhibits a strong negative optical density near 435 nm, indicative of a bandgap photo-bleaching (PB) signal, with a possible contribution from stimulated emission (Fig. 2b). Additionally, distinctive peaks of excited-state absorption (ESA) appear below 425 nm, resulting from the quantum confinement effect in our PeNCs[39–41] (Fig. S5).

A bi-exponential fit to the time profile of the first mode yields two lifetimes: 20.2 ps and 345.5 ps (Fig. 2c). The shorter component, decaying within tens of picoseconds, is typically associated with biexciton decay or the Auger recombination process[39,42–45]. However, given our exceptionally low pump energy of 0.5 nJ per pulse, this is better described by the charge carrier trapping process originating from shallow defects on the surface[46–48]. The intrinsic low power of our spectroscopic apparatus allows us to bypass the consideration of multi-exciton formation, which can occur when using high-power pump and requires many-body analyses to extract one-exciton dynamics from the complicated TA signals[49].

On the other hand, lifetimes on the order of hundreds of picoseconds are primarily associated with trap-mediated decay or recombination of charged excitons, known as trions[43,45]. It is important to note that the decay of unaffected excitons typically occurs on the order of a few nanoseconds[39,43]. The decay of charged excitons is observed even under low fluence, as their formation is closely tied to charge traps in surface defects[50,51]. Both observations are consistent with the high trap density of chloride-based perovskite nanomaterials, which create a significant barrier to their application as blue light emitters[50].

For the second mode, its time profile decays within one ps. Such sub-picosecond transition is often a complex signal and difficult to identify due to the concurrent contributions of coherent artifacts, such as scattering signals, within this temporal range[52]. The DAS of the second mode shows a negative peak at an energy level higher than the PB peak of the first mode, which is a characteristic of hot carrier thermalization and relaxation[44,53]. The pronounced ESA feature observed here is linked to the bandgap renormalization process[40,42]. Thus, we interpret the second mode as indicative of the initial excited state undergoing intra-band relaxation.

An overview of the discernible kinetics achieved through our technique is presented in Fig. 2d based on distinct lifetimes and spectral features. Note that the shorter decay of PB originates from the emptying of the conduction band-edge to shallow traps, while the longer PB decay is induced by the recombination of charge carriers to the ground state. The radiative decay lifetime is confirmed through time-resolved fluorescence measurements (Fig. S6).

Our unique contribution to the study of photo-dynamics is shown in the final panel of Fig. 2d, which will be discussed in the following sections. This portion illustrates the photoinduced substitution of the halide composition of PeNCs, occurring over an extended duration, spanning minutes. The figure is based on reports suggesting that the impact of halide composition primarily affects the valence band, driven by the interaction between hot phonons and holes[54]. Our AI-TA technique enables the observation and analysis of the distinct physical properties exhibited by each composition generated through light exposure.

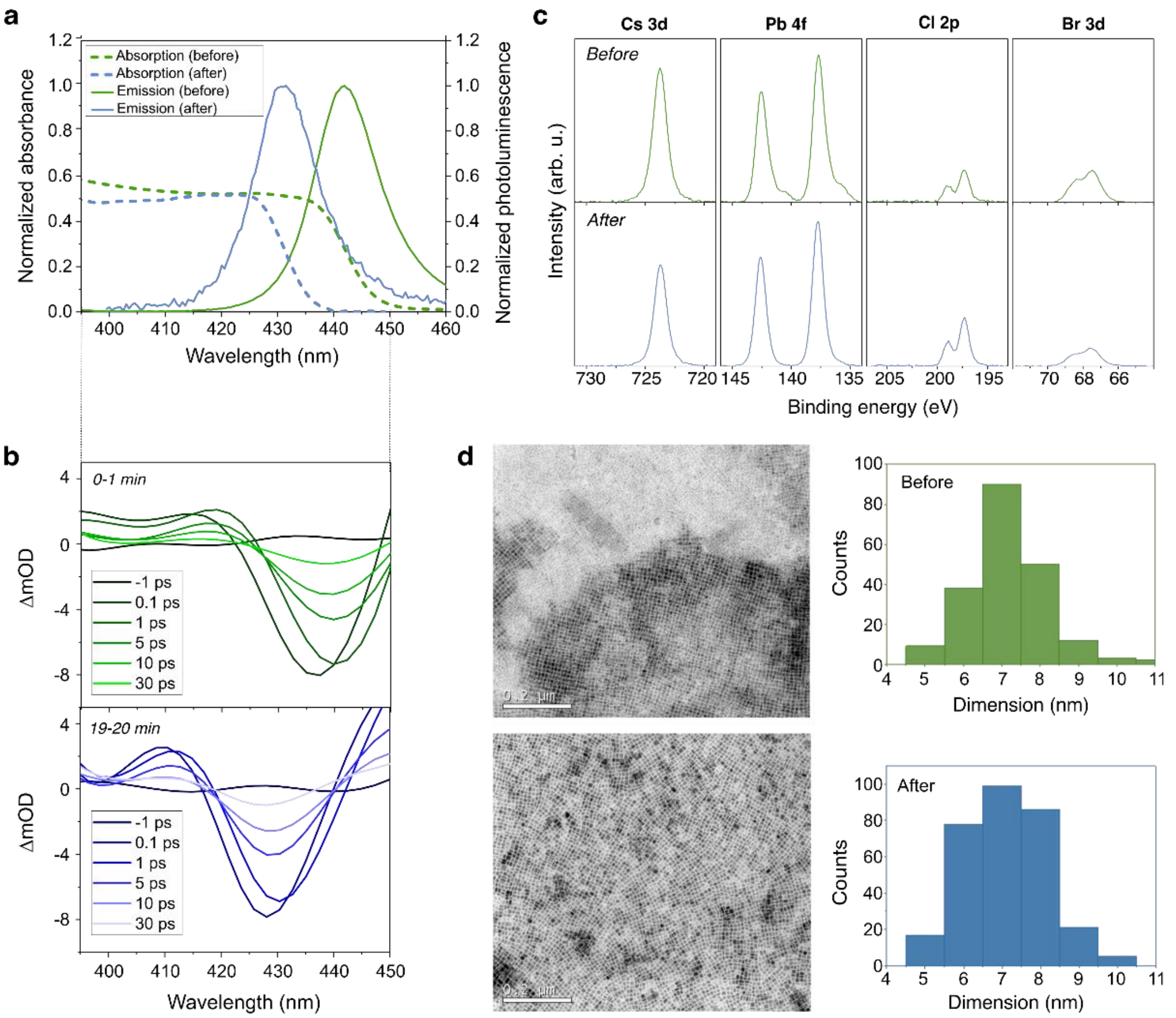

**Fig. 3. Photodynamics and photo-substitution of PeNCs.** (**a**) Comparison of absorption and emission spectra acquired before and after the AI-TA experiment of PeNC in chloroform. (**b**) Transient absorption maps of the first- and the last-minute scan of the experiment showing spectral evolution. The blueshift of the PB peak and corresponding steady-state spectra confirm the increased bandgap of the nanocrystal. (**c**) X-ray photoelectron spectroscopy (XPS) results for the PeNC sample before and after an AI-TA experiment of PeNC in chloroform. (**d**) TEM image of PeNCs measured before and after the AI-TA experiment of PeNC in chloroform and the histogram showing size distribution. The scale bar of the image of 0.2 μm.

## Photoinduced halide substitution of $CsPbX_3$ nanocrystal

The first subject of our study involves PeNCs with a mixed halide composition, specifically $CsPb(Br/Cl)_3$, dissolved in a chloroform/toluene mixture. Upon irradiation, the bromide ions in the PeNCs are substituted with chloride ions. As described earlier, this anion-exchange reaction occurs as a result of the photoinduced electron transfer process, in which the PeNCs extract halides from the surrounding solvents, leading to the formation of a new composition. The overall process can now be summarized by the following photochemical reaction:

$$\mathrm{CsPb(Br/Cl)_3} \xrightarrow{\mathrm{CHCl_3, h\nu\ (fs\ laser\ pulses)}} \mathrm{CsPb(Br\downarrow/Cl\uparrow)_3},$$

where PeNC, denoted as $\mathrm{CsPb(Br}\downarrow/\mathrm{Cl}\uparrow)_3$, differs from $\mathrm{CsPb(Br/Cl)_3}$ due to the substitution of Br atoms with Cl atoms.

Since the equilibrium between the nanocrystals and the solvated halide anions is reached almost instantly, within a few seconds[17], continuous DAQ under these conditions allows for *in situ* observation of the transforming PeNCs. The high-speed DAQ enabled by AI-TA significantly mitigates the influence of the pump and probe on the system during each scan. Note that each interferogram scan, including the data-saving process, takes less than 10 s, which is a substantial improvement over conventional technique. However, due to inherent noise in the data from a single interferogram, averaging is necessary (Figs. S7 and S8). The transient mapping shown afterward represents an average of a few scans, culminating in an approximate acquisition time of 1 min, which is required to delineate the transient features.

Figure 3a shows the absorption and emission spectra of PeNC in toluene/chloroform, before and after 20 min of AI-TA experimentation, and Figure 3b depicts spectral evolution of transient absorption spectra, reflecting the first and the last minute of the experiment. Both the absorption edge and the PB peak initially appeared around 440 nm but shifted to 430 nm, indicating a change in bandgap energy levels. The PB peak positions closely match the steady-state absorption and emission spectra measured before and after the experiment, as shown in the top panel of the figure. This measurement result provides evidence of a shifting bandgap while upholding the confined nature of the system.

To verify compositional changes, the remaining stock solution and the solution collected after the AI-TA experiment were analyzed using other materials analysis methods. X-ray photoelectron spectroscopy (XPS) analysis, shown in Fig. 3c (wide scan in Fig. S9), corroborates the compositional changes in the PeNCs before and after a separate 30-minute AI-TA experiment (Fig. S10). The peaks corresponding to the elements are magnified in the panel to highlight the compositional changes. The Cl/Br ratio shifts from 1.72:1.28 to 2.24:0.76 in the pre- and post-photo-substituted samples, respectively. Additionally, the size of the nanocrystals remains unchanged, suggesting that the observed shift in bandgap is not due to changes in the nanocrystal dimensions (Fig. 3d). The average sizes are 7.1 (± 1.1) nm and 7.2 (± 1.1) nm for the samples before and after photoinduced anion exchange reaction, respectively. The results confirm that halide exchange occurred while preserving most of the macroscopic physical properties, although some physical degradation may have also occurred (Fig. S11).

**Analysis of the parameters involved in halide substitution**

Before conducting an in-depth analysis of the AI-TA data scans, it is imperative to verify the orthogonality of the time parameters. The independence of these parameters is confirmed by monitoring the PB peak throughout the experiment, as depicted in Fig. 4a. The peak position of each successive data point falls within the error bars of both the preceding and following measurements, which validates the orthogonality between each transient absorption spectra map and the reaction time $t_{react}$. It is important to note that the reaction reaches completion after approximately 100 min, indicating a dynamic equilibrium between the halides in the PeNCs and those dissolved in the solvent. The photoreaction can also be characterized by an approximate lifetime of 46 min, as determined by fitting the peak position (Fig. S12).

The effects of the halide solvent and irradiation are validated through multiple control experiments. When toluene is used as the sole solvent, the peak position remains unchanged throughout the experiment (Fig. 4b). However, when chloroform is introduced to the solvent, the PB peaks exhibit a blueshift, highlighting the effect of solvent composition on the rate of photo-substitution within the PeNCs. The rate and final halide ratio of photoinduced exchange have been reported to depend on the ratio of toluene to halide solvents[22]. Additionally, the concentration of PeNCs, and consequently their optical density, is shown to influence the kinetic rates (Fig. S13a). The critical role of halide solvents is further demonstrated by the reverse substitution from chloride to bromide when dibromomethane is used as the halide solvent instead of chloroform (Fig. S13b).

The role of light as a reaction trigger is confirmed through a temporary pause in measurements (Fig. 4c). While the peak position shifts during continuous irradiation, it remains unchanged in the absence of light. The observations corroborate both halide solvents and irradiation as key factors in the photoconversion process between halides in PeNCs and those bound to the solvent.

Furthermore, the use of a high-repetition-rate oscillator may induce heating in addition to irradiation for photoinduced chemical reactions. To assess the potential contribution of this effect to the substitution process, we conducted a control experiment by heating a PeNC/chloroform solution in the dark. The steady-state spectra (Fig. S13c) remained unchanged, indicating that thermal effects are negligible.

Figure 4d presents processed data from the initial data segment, which is the averaged result of 10 interferogram scans and corresponds to less than one minute of $t_{react}$, where the nonlinear response is plotted with respect to $t_{pump\text{-}probe}$ and $\lambda_{probe}$. This yields information typically acquired from conventional transient absorption.

The inclusion of the $t_{react}$ axis emphasizes the advantage of our approach over conventional methods, as it enables the identification of variations in spectral or temporal signatures that depend on $t_{react}$, which in this case is associated with photo-substitution. Fig. 4e presents the decay profiles of the first mode, obtained from the SVD of data averaged over every ten scans, illustrating the relationship between $t_{pump\text{-}probe}$ and $t_{react}$. The plot clearly shows that the ultrafast rise of the PB signal, identified earlier as hot carrier relaxation, becomes shorter as reaction time progresses. The thermalization of hot carriers is known to depend on the excess energy provided by the pump above the bandgap[53]. Since our pump energy and fluence remain constant throughout the experiment, the observed behavior of the rise signal can be attributed to the bandgap energy increasing until it aligns with the pump range. It is important to note that this trend appears to plateau near 60 min, well before the PB peak shift halts. The early cessation of the change in thermalization rate may be due to the bandgap entering the pump spectrum range (Fig. S1).

From these results, we can infer that having excess energy results in a longer thermalization and bandgap renormalization process in the sub-picosecond decay. Also, the exponential fitting of PB decay shows a slight decrease in lifetime as chloride photo-substitution progresses (Fig. S14). Given that we have linked this parameter to the charge trapping of electrons in shallow traps, this relationship suggests that higher chloride content accelerates charge trapping. Interestingly, similar observations have been reported, where a faster charge-induced trapping rate was correlated with higher chloride content in PeNCs[46]. Since vacancies are more abundant in higher chloride compositions, the increased trap density facilitates better separation of charge carriers such as electrons. Moreover, because PeNCs are continually exchanging ions with surrounding solvent molecules, the high variability in trap density and, consequently, the lifetimes, can also be explained.

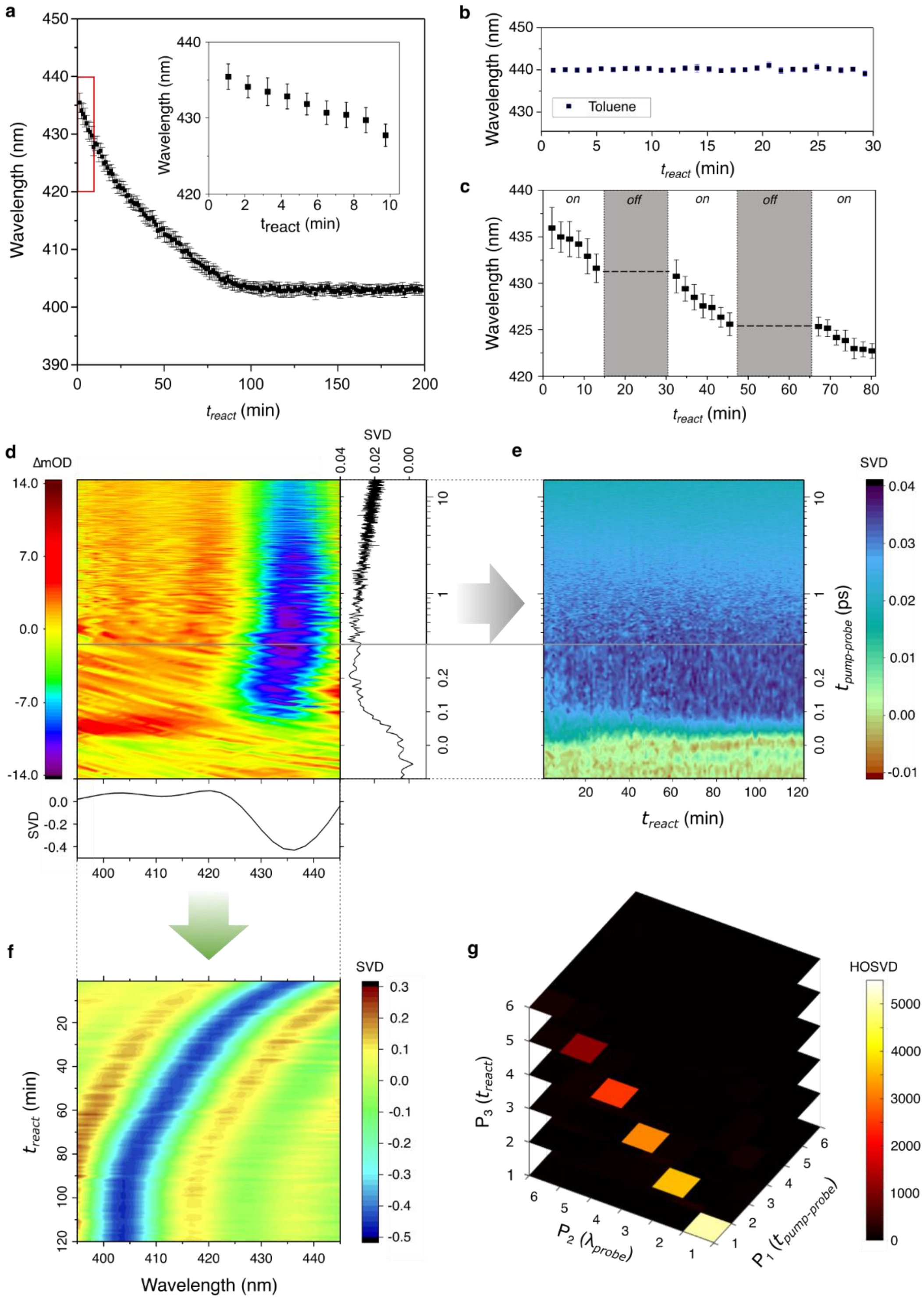

**Fig. 4. Comprehensive mapping of AI-TA analysis results.** (**a**) Extended peak tracking over three hours of the experiment. The inset zooms in to show that each sequential peak lies within the error bar of the prior measurement. (**b**) Analysis of the initial one-minute data acquired from AI-TA measurement. The trend along $t_{pump\text{-}probe}$ axis provides insight into the kinetic parameters and lifetime of the photophysical processes of PeNCs, while the $\lambda_{probe}$-dependent spectrum provides

spectral information of excited states. Along each axis, the first mode acquired from singular value decomposition is plotted. (**c**) Decay profiles of the first mode acquired from SVD of data segmented into one-minute intervals of $t_{react}$, showing its relationship with $t_{pump\text{-}probe}$. (**d**) $\lambda_{probe}$-$t_{react}$ correlation map showing the evolution of spectral characteristics with the duration of the experiment. (**e**) Reduced core tensor for extended two-hour-long measurement, which is the result of the high-order singular value decomposition (HOSVD) analysis of the AI-TA experiment data. Here, the index of each axis represents singular values arranged in descending order from the highest to the lowest, which are truncated to show only major modes involved in the dynamics.

In Fig. 4f, a similar analysis of the DAS of the first mode against $t_{react}$ shows that spectral characteristics are evidently influenced by the duration of the experiment. This $\lambda_{probe}$-$t_{react}$ correlation map reveals a blueshift in the photo-bleach peak, reaching 403 nm, consistent with the peak analysis shown in Fig. 3a. From the evidence presented so far, we can confidently assert that the real-time evolution of spectral characteristics is a result of the substitution of bromide with chloride. Overall, this illustrates the multidimensional nature of a single AI-TA measurement under photo-responsive conditions, a capability made possible by our technique.

Building upon this, we can parameterize three orthogonal variables based on the temporal order of the time segments defined earlier. First, $P_1$ represents the transient time-resolved response due to the incidence of the pump pulse, which is related to $t_{pump\text{-}probe}$. $P_2$ corresponds to bandgap energy-resolved information of PeNCs in terms of $\lambda_{probe}$, obtained from the FFT of the interferogram scan. Lastly, $P_3$ reflects the halide ratio of PeNCs, based on the unidirectional photo-substitution from bromide to chloride, and is associated with the total DAQ time divided into one-minute intervals corresponding to $t_{react}$. By examining the dynamic responses between any pair of these parameters, we can identify correlations within the system. Since the raw data arrays can be structured as a rank-three tensor using three-fold time-division multiplexing, we can isolate states or modes related to a single parameter of $P_n$ (n=1,2,3) and establish the relationships between these states through tensor decomposition analysis. In other words, the decomposed tensors can be projected into corresponding 2-dimensional planes.

To demonstrate and simplify the picture of parameter correlations, we apply higher-order SVD (HOSVD) to two-hour-long experimental results[55,56]. The truncated core tensor, or decomposed projection, of this analysis is portrayed in Fig. 4g (Table S1). The visualization summarizes and highlights the relationships discussed earlier. The prominent diagonal values along the $P_2$-$P_3$ basis correspond to the progression of different wavelength modes along the $t_{react}$ axis. In other words, it shows the shift in bandgap energy as the chloride composition increases and the bromide composition decreases. Conversely, the distinctly lower values, apart from the first mode of the $P_1$ basis, indicate that a single decay profile dominates, regardless of $\lambda_{probe}$ or $t_{react}$.

Through an iterative fabrication approach, often referred to as compositional engineering, remarkable strides have been achieved in developing efficient perovskite solar cells[57–60] and light-emitting devices[61]. This concept also extends to the quantum yield efficiency of PeNCs[54]. The importance of halide composition has been emphasized in studies highlighting the close link between the purity of a single photon source and its halide composition[62]. Halide exchange, usually induced by salt-based mechanisms, has been a topic of great interest for those who study the effects of compositional modifications while preserving morphology and minimizing other structural influences[42,46,53,63]. In particular, chloride substitution, despite its favorable bandgap for blue or ultraviolet light sources, has faced challenges in practical applications due to low quantum yield and relatively poor power conversion efficiency[50,64]. Although compositional engineering has proven effective, it remains a time-consuming and complex process. As a result, considerable efforts have been directed toward evaluating photoluminescence lifetimes through in-line synthesis[65]. Our strategy, utilizing AI-TA, offers a faster and more precise method for halide composition engineering.

### AI-TA measurement of photon-driven transformation process

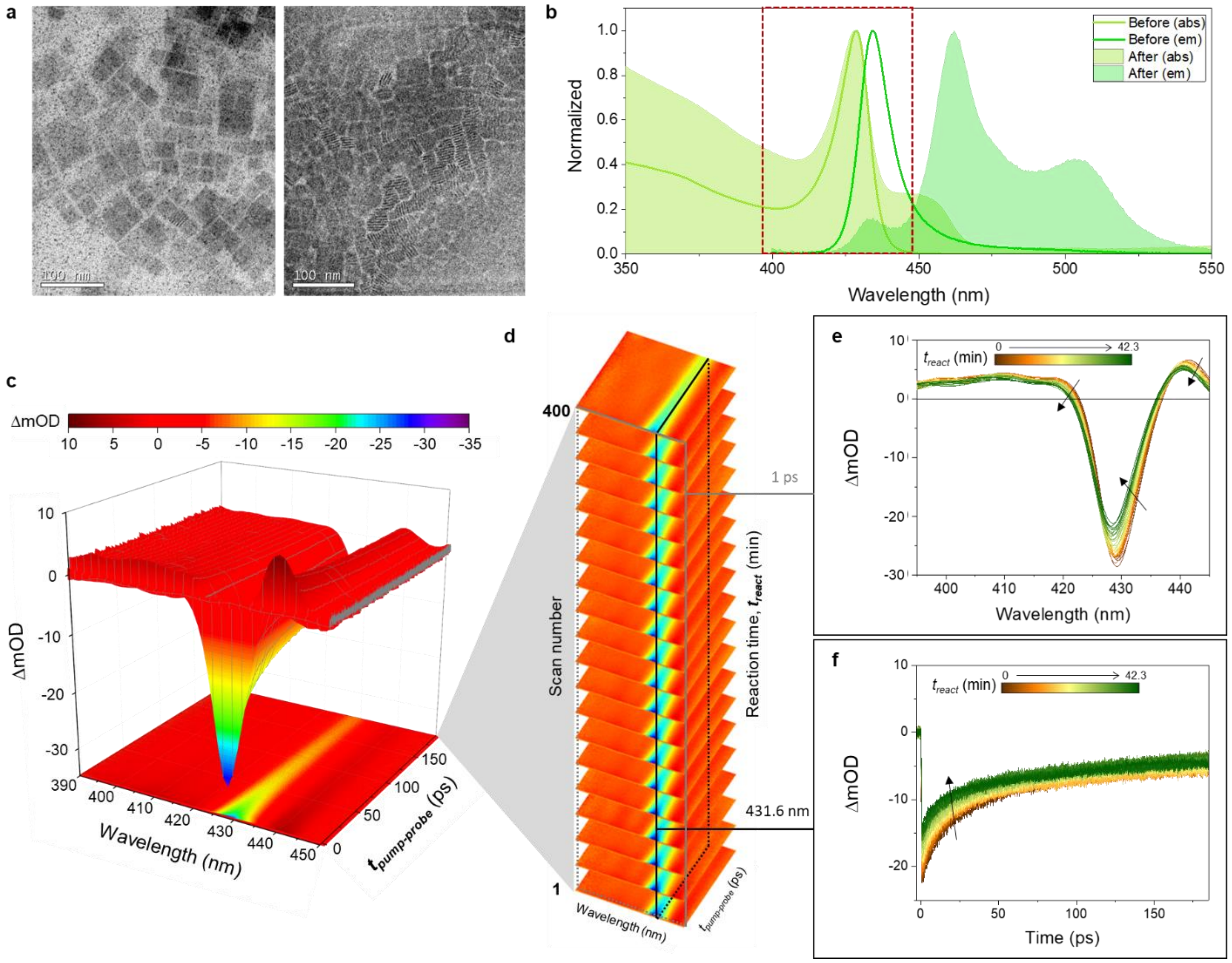

**Fig. 5. *In situ* and real-time analysis of 2ML PeNPLs kinetics using AI-TA.** (a) TEM image of PeNPLs standing face down and side up. The average size in unconfined dimensions is approximately 20 nm. (b) Normalized absorption and emission spectra of PeNPLs before and after the AI-TA experiment. The probe range is marked with a dotted red square. (c) A transient absorption profile of 2ML PeNPLs acquired by averaging 400 scans of AI-TA. (d) Projection by reaction time of individual scans participating in the averaged spectrum. (e) A collection of transient spectra at the pump-probe delay time of 1 ps for different reaction times. (f) A collection of time profiles at the probe wavelength of 431.6 nm for different reaction times.

In the next phase of our AI-TA investigation, we observed changes in the transient features of PeNPLs dispersed in hexane as the measurement progressed. PeNPLs are characterized by stronger exciton binding energy due to enhanced quantum confinement from thickness control. The thickness is quantified by the number of lead halide octahedra layers, referred to as monolayers (MLs), in the quantum-confined dimension. In addition to varying the halide composition mentioned earlier, the bandgap energy of PeNPLs can also be tuned by controlling the confinement thickness. This approach is effective for only few monolayers, as the effective Bohr radius for exciton in $CsPbBr_3$ is approximately 7 nm and the size of a cubic unit cell is 0.58 nm[12,30,66].

The composition investigated here consists of cesium lead bromide with 2 ML confinement. This is the thinnest variant, other than 1 ML, which consists of a single layer of lead halide octahedra without cesium cations and is stabilized with ligands. A TEM image confirms the formation of PeNPLs with an approximate size of 20 nm in the non-confined dimension (Figure 5a), which is known not to significantly affect the peak energy[67]. The absorption and emission spectra of the synthesized $CsPbBr_3$ PeNPLs show peaks near 430 nm, consistent with reported values, whereas bulk materials with the same composition show peaks near 530 nm[67,68]. The sharp excitonic peak confirms the large binding energy that is characteristic of PeNPLs.

After the AI-TA measurement, the steady-state spectra reveal the appearance of additional peaks in the longer wavelength region (Figure 5b). The redshifted spectrum matches the characteristic multiple peaks of higher-order PeNPLs with more MLs, indicating photon-driven transformation during the AI-TA measurement. Compared to photoluminescence, the higher-order peaks are less prominent in the absorption spectrum because they have higher quantum yields and therefore show stronger photoluminescence[24,67]. The transformation happening over the course of the experiment, visualized earlier in Fig. 1e, can be summarized with following reaction.

$$\text{2ML CsPbBr}_3\ (dispersed) \xrightarrow{\text{hν (fs laser pulses)}} \text{2ML CsPbBr}_3\ (agglomerate) \xrightarrow{\text{hν (fs laser pulses)}} 3\uparrow \text{ML CsPbBr}_3\ (aggregate)$$

Another important consideration in this experimental scheme is that the bandgap energies of 3ML and higher-order structures, including bulk material, are much lower and fall outside the spectral range of our near-ultraviolet probe[67]. Therefore, only the first half of the reaction involving only the 2ML and agglomerate structures are selectively detected. Also note that other perovskite phases, which can form in cesium-poor environments similar to PeNPLs, have higher bandgaps and are not considered in this investigation.

Typically, nonlinear responses are small, so transient absorption signals are usually averaged over extended DAQ periods. Here, it should be emphasized that the effect of sample coalescence is not very evident in the averaged TA measurement of PeNPLs shown in Figure 5c, which represents the collective response of 400 interferogram scans acquired over approximately 40 minutes of $t_{react}$. However, AI-TA enables us to analyze the data with fewer averages and varying $t_{react}$ durations (Figure 5d). Unlike traditional TA spectroscopy, we can analyze the spectrum while adjusting the desired measurement time.

By separately analyzing the data in one-minute segments of $t_{react}$, we observe a blueshift in the transient signals. Additionally, the peak intensity decreases, likely due to sample degradation, as similar trends are observed in the AI-TA scan of PeNCs in toluene (Figure S11), where the intensity diminishes over the course of the scans. Such sample changes would likely interfere with time-profile measurements in conventional systems, where TA spectra are averaged over fixed time intervals.

The complete disappearance of AI-TA signal, marking the complete consumption of 2ML variants, takes over 2 hours. However, in this study, extended measurements after 400 scans are affected by aggregation at the focus and the formation of higher-order structures, and thus these data are discarded from the analysis (Fig. S15). Consequently, this result cannot be directly compared with the post-experiment steady-state spectra. TEM imaging of the measurement sample after collection reveals some mosaic-like structures of PeNPLs (Fig. S16), consistent with previously reported observations.

Spectra of transient signals at a fixed $t_{pump\text{-}probe}$ of 1 ps (Figure 5e) and a fixed probe wavelength of 431.6 nm (Figure 5f) were collected to compare the evolution of the signals over reaction time, revealing clear and definite changes. The shifts in the bandgap bleach signals to lower wavelengths are subtle but noticeable in the TA spectrum. The bandgap bleach peak shifts by approximately 1 nm, accompanied by changes in the distribution of excited carrier populations, as indicated by shifts in both sidebands (Figure S17). This effect becomes more pronounced when comparing normalized spectra acquired at different reaction times. This observation can be attributed to the preferential degradation of PeNPLs with larger non-confined dimensions. However, the impact of this parameter on the shift of the excitonic absorption peak is less significant compared to the number of layers, which serves as a more discrete parameter[67] (Figure S18).

The collection of time profiles at a fixed wavelength (Figure 5f) show that the decay trend also changes. The triexponential fit of the time profiles yields three different time constants (Fig. S19), which can be interpreted in terms of kinetics similar to those identified earlier in Figure 2, representing hot carrier thermalization, charge trapping, and carrier recombination[69].

The shortest component, corresponding to the hot carrier thermalization process, provides a unique insight into the dynamics of PeNPLs. The spectral and temporal signature comprises a subpicosecond peak redshift, appearing as a decay at shorter wavelengths and as a rise at longer wavelengths. The excess energy of the hot state is lost via electron-phonon coupling. As the measurement progresses, the hot carrier relaxation lifetime of PeNPL gradually increases from approximately 0.08 ps to 0.12 ps, indicating a bottleneck in the relaxation dynamics.

Although the hot phonon bottleneck is reported to have a negligible impact on highly confined $CsPbBr_3$ PeNCs[70], such a slowdown in hot carrier cooling has been observed in superlattice PeNPLs[71]. Similarly, the efficiency of hot phonon cooling is much better in colloidal samples compared to film counterparts with the same number of MLs[72]. In both cases, carrier cooling has been compared between colloidally dispersed and dried samples and confirmed through power dependence measurements. Another study on a sample aged for 80 days and measured with time-resolved upconversion spectroscopy also reports an increased lifetime[32].

Likewise, our results can be interpreted as an agglomeration-induced suppression of hot carrier relaxation. The excess energy of the hot state is dissipated via electron-phonon coupling to form longitudinal optical phonons[71,73]. The thermal energy is subsequently transformed into out-of-plane modes and dissipated into the environment, particularly efficiently in highly confined small PeNCs and PeNPLs. However, the formation of stacked structures, known as superlattices, suppresses this out-of-plane coupling and causes a bottleneck in thermal relaxation. In our case, the observation of the hot phonon bottleneck was achieved not by measuring power dependence, but through fast DAQ of progressing photoinduced agglomeration of PeNPLs.

The other lifetime, associated with the charge-trapping process, remains almost consistent. This lifetime is assumed to be intrinsic to the state of PeNPLs undergoing AI-TA measurement, where femtosecond laser pulses create a photochemically dynamic and highly stressed environment. Another interesting observation is that the lifetime of the longest components also shows elongation. This component reflects exciton recombination, which may involve trapped charges, as observed in the case of PeNCs. This trend could be related to the size reduction observed alongside the PB peak shift, which involves the removal of incomplete unit cells with vacancies at the edges of the nanoplatelets.

This trend can also be visualized in collections of highly averaged and normalized time profiles at longer wavelengths. One advantage of AI-TA is that it allows control of the length of data acquisition for averaging, enabling us to validate the minute changes observed here. Shorter acquisition times result in a lower signal-to-noise ratio, so integrating longer reaction times and analyzing the data over these periods helps clarify the trend observed in the experiment. This is demonstrated by averaging over different reaction times. Comparing the averaged results for 1 minute, 5 minutes, and 10 minutes of $t_{react}$ (Figure S20) confirms the observations made with one-minute averages. Again, we emphasize the strength of the AI-TA technique, particularly its ability to account for changes occurring throughout the measurement process.

**Conclusion and outlook**

In this study, we successfully applied the AI-TA technique to the *in situ* observation of the photophysical behavior of perovskite nanomaterials undergoing photoinduced chemical reactions and transformation processes. We investigated specialized forms of light-induced photodegradation and demonstrated that, in some cases, the measurement process itself can influence the results of a spectroscopic experiment. Our AI-TA technique helps elucidating the impact of such measurement effects. This method offers a streamlined approach for investigating photoinduced dynamic processes across a broad temporal range, from femtoseconds to minutes. By conceptualizing the time taken for the experiment as a chemical reaction axis, this technique can be regarded as a new form of multidimensional spectroscopy. While this experiment focused on perovskite nanomaterials, the methodology holds promise for extension to other photoreactive chemical or biological systems, thereby expanding the scope of its applicability.

In summary, the first study presented here enhances our understanding of the relationship between trap states and chloride-to-bromide variation in $CsPb(Br/Cl)_3$ PeNCs as a function of reaction time and provides an effective strategy for investigating similar systems. It is worth noting that both experimental verification and theoretical calculations of halide composition variations would typically require substantial resources and time investment. The AI-TA technique effectively addresses these challenges by reducing several months of investigation to just a few minutes, while requiring only a small amount of material for multiple compositions. The second study involving PeNPLs represents an interesting example of bridging photophysical observations between films and colloidal nanoplatelets. This study offers a deeper understanding of the excited state dynamics and spectral shifts that occur as colloidally dispersed particles aggregate. With the emergence of advanced data handling technologies, where the acquisition of extensive datasets for analysis is crucial, our work stands to make a meaningful contribution to the future of materials research.

## Methods

### Asynchronous and interferometric transient absorption (AI-TA) spectrometer

The outputs of second-harmonic generation from two Ti:Sapphire mode-locked lasers (Rainbow2, Femtolasers) were used as fundamental light sources, serving as the pump and probe lasers, as well as the radiation sources for photo-substitution reactions of PeNCs. Their repetition rates were stabilized to 80.0 MHz for the probe and 80.0 MHz + $\Delta f$ for the pump by phase-locking to an 8-channel frequency synthesizer (HS9008B, Holtzworth) referenced by a GPS-disciplined Rb atomic clock. The detuning frequency, $\Delta f$, was set to 51.2 Hz, 153.6 Hz, or 460.8 Hz (8 fs, 24 fs, or 72 fs time steps respectively) depending on the measurement window of pump-probe time. The digitizer was operated at 80.0 MHz, synchronized with the frequency of the probe pulses, to record light intensity within a preset time interval. BBO crystals (Castech) with different critical phase matching conditions were used to convert the laser source wavelength from near-infrared to near-ultraviolet region. The second harmonic generation signal, produced by the pump and probe overlap on a nonlinear crystal, was used as an optical trigger that established the starting point of decay profiles. Pump and probe pulses were focused using UV-enhanced off-axis parabolic mirrors at a 1 mm sample cell (21/Q/1, Starna Scientific). The cell contained a sample with an optical density of approximately 0.3 and a volume of 0.4 mL and was vigorously stirred using a magnetic bar. This stirring serves the dual purpose of minimizing the photodegradation and the photothermal signals while expediting the equilibrium for anion exchanges, which occur rapidly enough to be considered simultaneous[17]. After passing through the sample, the interferogram of the probe pulse was acquired using a balanced photodetector (PDB415A, Thorlabs) in a symmetric configuration[74]. Every experiment was conducted under ambient, room-temperature conditions. MATLAB with embedded function packages was used to analyze data, including the fast Fourier transform. A more detailed instrumental and theoretical description of the method is provided in Supplementary Notes.

**Chemicals**

Lead acetate trihydrate ($Pb(CH_3COO)_2.3H_2O$, 99.99%), cesium carbonate ($Cs_2CO_3$, reagent Plus, 99%), lead(II) bromide (PbBr2, ≥98%), benzoyl bromide ($C_6H_5COBr$, 97%), benzoyl chloride ($C_6H_5COCl$, 98%), acetone ($CH_3COCH_3$, HPLC, ≥99.9%), toluene (anhydrous, 99.5%), hexane ($C_6H_{14}$, ≥99%), chloroform (HPLC Plus, ≥99.9%), octadecene (ODE, technical grade, 90%), oleylamine (OLAM, 70%), and oleic acid (OA, 90%) were purchased from Sigma-Aldrich. All chemicals were used without any further purification.

**Synthesis**

**$CsPbX_3$ (X=Cl, Br) perovskite nanocrystals** $CsPbX_3$ (X=Cl, Br) PeNCs were prepared by following the reported procedure[75] with slight modifications. In a typical synthesis, cesium carbonate (16 mg), lead acetate trihydrate (76 mg), 0.3 mL of OA, 1 mL of OLAM, and 5 mL of ODE were loaded into a 25 mL 3-neck round-bottom flask and degassed under vacuum for 1 h at 100 ℃. Subsequently, the temperature increased to 170 ℃ under Ar, and 0.6 mmol of the mixture of benzoyl chloride/bromide (a precursor ratio of 1:1) was swiftly injected. The reaction mixture was immediately cooled down in an ice–water bath. Finally, 5 mL of toluene was added to the crude NC solutions, and the resulting mixture was centrifuged for 10 minutes at 3500 rpm. The supernatant was discarded, and the precipitate was redispersed in 5 mL toluene for further use.

**2ML $CsPbBr_3$ perovskite nanoplatelets** All-inorganic PeNPLs with the composition $CsPbBr_3$ were synthesized using a previously reported ligand-assisted reprecipitation method under ambient conditions[67]. Two precursors were prepared prior to conducting the synthesis. One was a solubilized lead bromide (0.01 mmol) in a mixture of 0.1 mL of oleylamine, 0.1 mL of oleic acid, and 10 mL of toluene. The mixture should be stirred and heated at 100 ℃ until $PbBr_2$ powder is fully dissolved. The cesium-oleate precursor was prepared by dissolving 0.1 mmol $Cs_2CO_3$ in 10 mL oleic acid. This was also stirred and heated at 100 ℃ until fully dissolved. The ratio of precursors to antisolvent in the synthesis determines the thickness of the synthesized PeNPLs. In the case of 2ML PeNPLs, 0.15 mL of Cs-oleate precursor was added to vigorously stirring 3 mL of $PbBr_2$ precursor. Immediately, 2 mL of acetone serving as an antisolvent was injected and the mixture was kept stirred for 1 minute. Then the solution was transferred to a 50 mL conical tube and centrifuged at 4000 rpm for 3 minutes. The supernatant was discarded, and the precipitate was redispersed in 2 mL hexane. All PeNPL solutions used for AI-TA experiments were kept for no longer than two weeks after synthesis.

**Steady-state absorption and emission spectroscopy**
The absorption and emission spectra of nanocrystals were acquired using Lambda 465 (Perkin Elmer) and Duetta (Horiba), respectively. A quartz cell with a path length of 10 mm was used for the measurement, and the sample was diluted to one-tenth of the stock solution.

**X-ray diffraction**
A Rigaku D/Max Ultima III X-ray diffractometer with graphite-monochromatized Cu Kα ($\lambda$ = 1.54056 Å) radiation was used to identify the crystal structure of PeNCs (Fig. S21). The PeNC sample solution was drop-casted onto the substrate for measurement. The diffractometer was set to 40 kV with a current of 30 mA, and the spectra were collected from 20° to 60° with a 0.01° sampling width.

**X-ray photoelectron spectroscopy**
K-alpha (Thermo VG, U.K.) instrument was used to measure the XPS spectra of the PeNCs. The PeNC samples were spin-coated on the silicon substrate for the measurement. The wide scan and narrow scan spectra were obtained by using a monochromatic Al X-ray source (1486.6 eV). The energy step sizes of the survey and narrow scans were 1 eV and 0.1 eV, respectively. The binding energy of the measurement was calibrated by the standard electron energy of 284.8 eV corresponding to the photoelectron from the carbon 1s state.

**Transmission electron microscopy**
The estimation of PeNC size was performed using a Tecnai G2 F30ST (FEI) microscope. For high-resolution measurement, a 300 S/TEM (Double Cs Corrected Titan3 G2 60, FEI) equipped with ChemiSTEM technology is used.

**Time-correlated single photon counting**
TCSPC measurements were conducted using FluoTime300 (PicoQuant). The sample was excited with a laser with a center wavelength of 405 nm. The photon count data was binned at 25 ps. For the stock sample, PeNC in toluene, and the PeNC sample after the experiment, the detection wavelength was set to 439.8 nm (5 nm bandpass) and 419.8 nm (5 nm bandpass), respectively.

**Acknowledgments**
We wish to thank K.J. Lee and J.M. Lim for their discussion and comments on the project, B.G. Choi for the assistance with the debugging of scripts, D.W. Jeong for insightful comments on materials analysis, W.J. Cho for advice on synthesis protocol, and RIAM (Research Institute of Advanced Materials, Seoul National University) for TCSPC measurement.
**Funding:** This work was supported by the Institute for Basic Science (IBS-R023-D1).
**Author contributions:** Conceptualization: G.R.H., T.H.Y., M.C.; Funding acquisition: M.C.; Investigation: G.R.H. – AI-TA/UV-VIS/PL/TCSPC, M.N.A. – XRD/XPS/TEM; Methodology: G.R.H., T.H.Y., M.C.; Project administration: G.R.H., T.H.Y., M.C.; Resources: M.N.A., G.R.H., K.S.J. - materials, N.S.H., J.W.K.,G.R.H. – instrumentation; Software: J.W.K., G.R.H., H.J.; Supervision: T.H.Y., M.C.; Validation: G.R.H., T.H.Y.,H,J., M.C.; Visualization: G.R.H., T.H.Y., M.C.; Writing - original draft: G.R.H.; Writing - review & editing: All authors.
**Competing interests:** Authors declare that they have no competing interests.
**Data and materials availability:** All data are available in the main text or the supplementary materials.

Supplementary Information for

# *In situ* and real-time ultrafast spectroscopy of photoinduced reactions in perovskite nanomaterials

Gi Rim Han[1], Mai Ngoc An[1,a], Hyunmin Jang[1,2], Noh Soo Han[1,b,] , JunWoo Kim[1,c], Kwang Seob Jeong[1,2], Tai Hyun Yoon[1,3,]*, Minhaeng Cho[1,2,]*

[1]Center for Molecular Spectroscopy and Dynamics, Institute for Basic Science (IBS); Seoul, 02841, Republic of Korea.

[2]Department of Chemistry, Korea University; Seoul, 02841, Republic of Korea.

[3]Department of Physics, Korea University, Seoul, 02841, Republic of Korea.

*Corresponding authors. Email: mcho@korea.ac.kr (MC) or thyoon@korea.ac.kr (THY)

[a]Present address: Quantum Science Ltd. TechSpace One, Sci-Tech Daresbury, Keckwick Lane, Warrington, WA4 4AB, United Kingdom.

[b]Present Address: Korea Research Institute of Standards and Science, Daejeon, 34113, Republic of Korea.

[c]Present Address: Department of Chemistry, Chungju, Chungcheongbuk-do, 28864, Republic of Korea.

Corresponding author: mcho@korea.ac.kr (MC) or thyoon@korea.ac.kr (THY)

**The PDF file includes:**

Supplementary Notes S1 to S4

Figs. S1 to S21

Table S1

Reference 76

## Supplementary Notes

### Supplementary Note S1. AI-TA setup

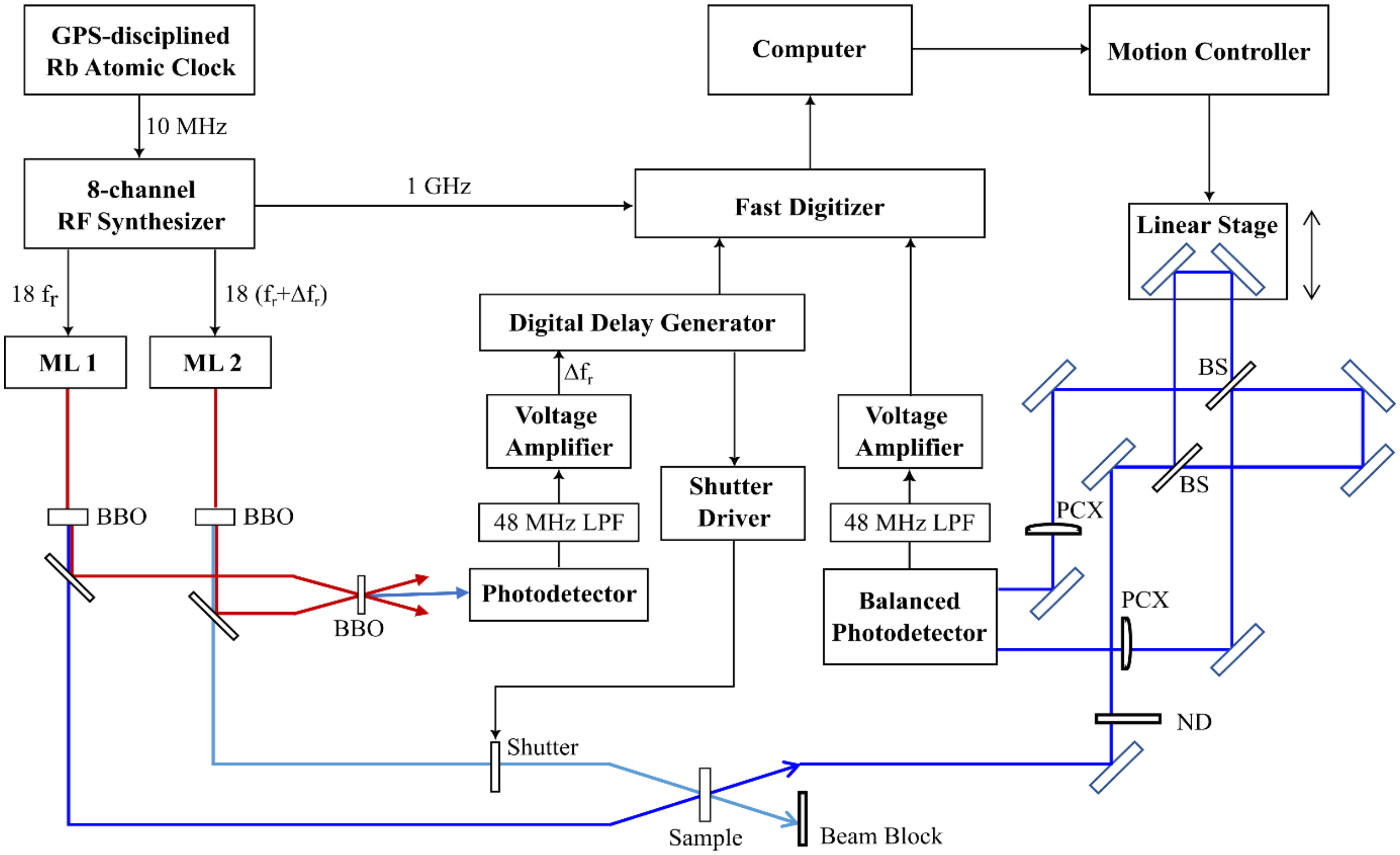

In our AI-TA experimental setup, we employ two Ti:Sapphire mode-locked lasers, ML1 and ML2 (Rainbow2, Femtolasers), as the fundamental laser sources. Both lasers deliver pulses with a duration of approximately 7 fs and a full-width-at-tenth maximum (FWTM) of over 330 nm. The repetition frequencies of ML1 and ML2 are set to 80 MHz + $\Delta f_r$ and 80 MHz, respectively, where $\Delta f_r$ is the detuning frequency.

For perovskite nanocrystal (PeNC) measurements, $\Delta f_r$ is adjusted to 51.2 Hz ($\Delta t$ = 8 fs) to capture an ultrafast decay pattern or to 409.6 Hz ($\Delta t$ = 72 fs) for extended measurements spanning up to nanoseconds. In the case of perovskite nanoplatelet (PeNPL) measurements, $\Delta f_r$ is set to 153.6 Hz, enabling pump-probe time delay scanning at intervals of 24 fs.

The repetition frequencies of both lasers are phase-locked to reference frequencies at the 18$^{th}$ harmonic (1.44 GHz) of their fundamental frequencies. These reference frequencies are independently generated from an 8-channel frequency synthesizer (Holzworth, HS9008B) whose time base is phase-locked to a 10 MHz reference signal from a GPS-disciplined Rb atomic clock.

During the experiment, feedback signals controlling the cavity lengths of the mode-locked lasers, via their piezo actuators, stabilize the repetition rates with a precision of mHz at 1-s integration time. The sampling rate is synchronized with the probe laser repetition rate (80.0 MHz) to ensure signal acquisition from every probe pulse. The system achieves a maximum time scan range of up to 12.5 ns.

BBO crystals (Castech) are used for the second-harmonic generation, converting the laser source wavelength from the near-infrared region (750-900 nm) to the near-ultraviolet region (390-450 nm). The phase matching angles and thickness of BBO crystals are $\phi$ = 29.2° and $d$=0.2 mm for the pump, and $\phi$=27.5° and $d$=0.02 mm for the probe.

The pump beam has a narrower spectral bandwidth of 13 nm full-width at half-maximum (FWHM) and a power of approximately 0.5 nJ/pulse, while the probe beam exhibits a broader spectral bandwidth of 36 nm FWHM and 0.025 nJ/pulse. Both beams are focused onto the sample using a UV-enhanced aluminum-coated off-axis parabolic mirror with a focal length of 101.6 mm. The pump diameter at the probe focus is approximately 75 μm, corresponding to a fluence of 11.3 $\mu J/cm^2$ per pulse.

At the focal position, a rectangular quartz cell (21/Q/1, Starna scientific) with a 1 mm pathlength is used, and the solution inside is continuously stirred with a magnetic bar to minimize thermal effects and maintain a rapid equilibrium of halide species between the nanocrystals and the solvent. Using Gaussian beam estimation, the irradiated volume at the focal spot is calculated to be on the order of $10^{-6}$ mL. The total sample volume used in the experiment is approximately 0.5 mL. The optical density of the sample at the pump wavelength is maintained at 0.4 – 0.6 for PeNC experiments and 0.1 – 0.2 for PeNPL experiments.

After passing through the sample, the pump beam is blocked, and the probe is directed to a Mach-Zehnder interferometer to generate an interferogram. Unlike our group's previously reported interferometric ASOPS work[3], the probe field in this setup is self-referenced using the homodyne detection method with a balanced photodetector[39]. Although this approach does not allow for the simultaneous acquisition of refraction spectra, this symmetric detection[53] offers a significant advantage in achieving a high S/N ratio for detecting the TA signal even in the presence of scattering processes. This improvement arises because the method ensures identical phase fronts for the interfering beams, enhancing the visibility of the interferogram.

In the self-referenced interferometer, the probe beam is split into two paths, one of which is directed to a linear stage (XMS50-S). Using a linear stage motion controller (XPS-RL), the pathlength of this arm is precisely controlled to generate a probe-to-probe time delay that results in an interference signal. After passing through the Mach-Zehnder interferometer, the recombined outputs are measured using a balanced photodetector (PDB415A, Thorlabs). This setup effectively cancels out common environmental noise and low-frequency signals. The scanning speed of the stage is critical to meet the Nyquist sampling criterion, ensuring accurate performance of the fast Fourier transform.

The scanning stage speed is maintained at 2.0 μm/s (or 2.3 μm/s) for continuous scanning, as the fastest oscillation component to be detected corresponds to a 390 nm wavelength with a period of approximately 1.3 fs. The chosen speed for the linear delay stage should be slow enough not to affect the probe time profile or the intensity modulations due to the pump pulses. Each interferogram scan, including the data-saving process, was completed in approximately 6 s, which is significantly more efficient than traditional grating-based spectral measurement techniques.

To generate the trigger signal, portions of the fundamental beams from ML1 and ML2 are extracted using dichroic mirrors. Silver parabolic mirrors with a focal length of 25.4 mm are used to focus the two parallel beams onto individual nonlinear crystals. A 1 mm-thick BBO crystal ($\phi$ =29.2°, Eksma optics) is employed to produce a sum-frequency-generation (SFG) signal, which is detected by a silicon detector (DET10A, Thorlabs). This trigger signal establishes the zero point of the decay profile, corresponding to the temporal and spatial overlap of the pump and probe pulses.

The detected signals are then amplified by 50 dB using a voltage amplifier (DHPVA−101, Femto) and then routed to the external trigger port of a digital delay generator (DG645, Stanford Research Systems). When the signal voltage exceeds a predefined threshold, the delay generator produces a 5 V TTL signal, which is then fed into the digitizer to initiate the data acquisition sequence. The delay generator's output signal lasts for 19.5 ms and could be "rearmed" by subsequent trigger signals after this period. This setup allows skipping some trigger signals to reduce the sampling rate under conditions of high detuning frequencies. Each trigger defines the time zero point for subsequent data acquisition, ensuring precise synchronization for averaging.

The digital delay generator is also connected to an optical shutter system positioned in the pump beam path before the sample. This shutter is controlled by a four-channel shutter driver (SR474, Stanford Research Systems) coupled to the digital delay generator. The shutter (SR475, Stanford Research Systems) opens only during designated intervals, effectively minimizing photodegradation caused by prolonged exposure to pump pulses.

Both the probe and trigger signals are passed through a 48 MHz low-pass filter to remove the 80 MHz component and its harmonics. The probe signal is then amplified by 20 dB using a voltage amplifier (DHPVA-101, Femto) and digitized using a fast digitizer (Spectrum, M3i.4861), which is synchronized to a 1 GHz reference signal from the multichannel RF synthesizer. The sampling rate is set to 80 MHz, matching the probe repetition frequency to ensure that data recording is synchronized with the probe beam repetition rate. Using the trigger signal, time-tagged points of the interferogram are collected and processed with a fast Fourier transform (FFT), yielding the effective signal spectrum.

## Supplementary Note S2. Theoretical description of AI-TA spectroscopy

The detailed theoretical descriptions of homodyne AI-TA in the NIR region and heterodyne AI-TA in the NIR (pump) and NUV (probe) regions were provided in our previous papers[3,4,55]. However, since the description of self-referenced (homodyne) asynchronous and interferometric TA has not been presented before[39], here we briefly reviews the fundamental theory, focusing only on the key differences from the previous works.

Transient absorption spectroscopy detects third-order nonlinear responses generated by the interaction of the pump and probe pulses. The signal is proportional to the third-order polarization $P^{(3)}$, which acts as an oscillating source for the generation of the signal electric field. Within the slowly varying envelope approximation, we have

$$E^{(3)}(\boldsymbol{k},t) \propto iP^{(3)}(\boldsymbol{k},t),$$

where $\boldsymbol{k}$ is the wave vector of the probe pulse.

Since light is detected as intensity to the photodetector, we observe the square of the electric field. For signals arising from pump- and probe-resonant transition, the intensity of the homodyne-detected signal field is given as:

$$\begin{aligned} I(t) &= \left|\left(E(\boldsymbol{k},t) + E^{(3)}(\boldsymbol{k},t)\right) + \left(E(\boldsymbol{k},t-\tau) + E^{(3)}(\boldsymbol{k},t-\tau)\right)\right|^2 \\ &= |E(\boldsymbol{k},t)|^2 + \left|E^{(3)}(\boldsymbol{k},t)\right|^2 + |E(\boldsymbol{k},t-\tau)|^2 + \left|E^{(3)}(\boldsymbol{k},t-\tau)\right|^2 \\ &+2Re\left[E^*(\boldsymbol{k},t)E^{(3)}(\boldsymbol{k},t)\right] + 2Re\left[E^*(\boldsymbol{k},t-\tau)E^{(3)}(\boldsymbol{k},t-\tau)\right] \\ &+2Re[E^*(\boldsymbol{k},t)E(\boldsymbol{k},t-\tau)] + 2Re\left[E^*(\boldsymbol{k},t-\tau)E^{(3)}(\boldsymbol{k},t)\right] \\ &+2Re\left[E^*(\boldsymbol{k},t-\tau)E^{(3)}(\boldsymbol{k},t)\right] + 2Re\left[E^{(3)}(\boldsymbol{k},t)E^{(3)}(\boldsymbol{k},t-\tau)\right] \end{aligned}$$

Through the Fourier transformation of the above signal intensity with respect to $\tau$, only the interference terms are selectively measured. Note that the term $2Re[E^*(\boldsymbol{k},t)E(\boldsymbol{k},t-\tau)]$ is just the autocorrelation of the incident probe beam and because $|E(\boldsymbol{k},t)| \gg \left|E^{(3)}(\boldsymbol{k},t)\right|$, $2Re\left[E^{(3)}(\boldsymbol{k},t)E^{(3)}(\boldsymbol{k},t-\tau)\right]$ is ignored. Then, the remaining $\tau$-dependent interference signal $S(t,\tau)$ is given as follows:

$$S(t,\tau) \propto 2Re\left[E^*(\boldsymbol{k},t)E^{(3)}(\boldsymbol{k},t-\tau)\right] + 2Re\left[E^*(\boldsymbol{k},t-\tau)E^{(3)}(\boldsymbol{k},t)\right]$$

In the previous description of the heterodyne detection scheme, only the second term is involved in the measured signal. Therefore, the signal obtained by adopting the self-referenced interference detection method is stronger than the heterodyne method. The Fourier transform with respect to $\tau$ yields frequency-dependent information. In addition, the laboratory time $t$ can be converted into the pump-probe effective time, using the down conversion factor, which yields the

time- and frequency-resolved TA spectra, $S(t_{eff}, \omega)$. Then, the frequency-dependent TA signal is divided by the steady-state signal before the pump can be expressed as:

$$\frac{S(t_{eff}, \omega)}{S(t_{eff} < t_{trigger}, \omega)} = \exp[-\Delta\alpha(t_{eff}, \omega)]$$

where $\Delta\alpha$ is the transient absorption spectrum. Note that the pump incidence time is defined by the trigger signal, and the starting point of the effective time, or 'time zero,' is described in relation to the defined trigger time, $t_{trigger}$.

## Supplementary Note 3. Correction of group velocity dispersion effect on AI-TA spectrum

Due to the broad bandwidth of the probe pulse and the presence of transmitting optics in the paths of both the pump and probe beams, group velocity dispersion (GVD) occurs. This results in a chirped response of the nonlinear signals, causing the starting point of transient signals to vary across different spectral components, as shown in the figure below (left). Correcting for GVD is crucial in ultrafast spectroscopy to achieve accurate temporal resolution and to properly analyze sub-picosecond responses. In this setup, the chirp of the probe pulse is corrected using a conventional fitting method[50]. The corrected spectrum is shown in the image on the right.

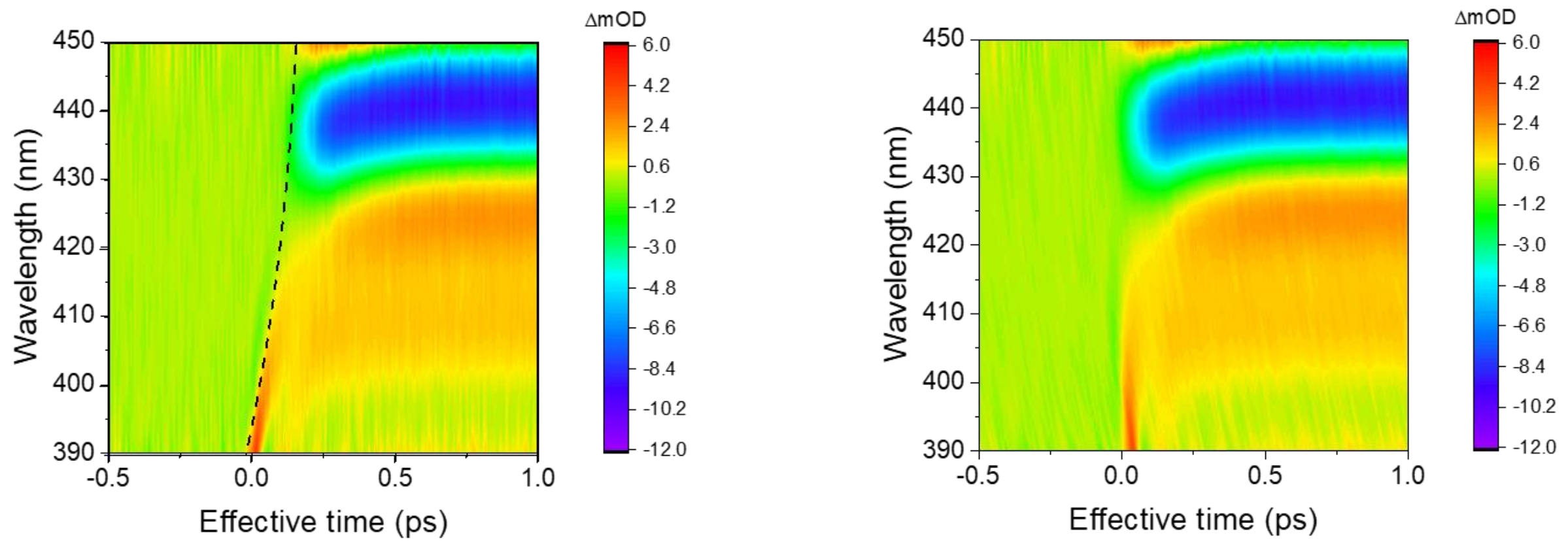

**Supplementary Note 4. Kinetics modeling of the photoinduced substitution**

Various chemical reactions following the primary photoreaction severely complicate the description of the kinetics of photo-bleach (PB) peak shift during AI-TA experiments. However, it is possible to construct an elementary dynamics model for the overall process under the assumption that the bandgap energy is proportional to the chloride-to-bromide anion ratio. Based on this assumption, the PB peak energy (or the bandgap energy) for each composition can be approximated as a variable that reflects the concentration of chloride anions within the PeNC lattice.

This description of the PB shift is associated with only the anion species within the PeNC and does not include contributions from free anions or radicals, such as $\mathrm{Cl^-}$ , $\mathrm{Br^-}$, and $\mathrm{CHCl_2}\cdot$, as these species are not detected by our probe pulses.

To begin, we can identify three major kinetic processes contributing to the peak shift dynamics:

$$\mathrm{PeNC(Cs_{\mathit{n}}Pb_{\mathit{n}}Cl_{\mathit{m}}Br_{3\mathit{n}-\mathit{m}})} + h\nu \longrightarrow \mathrm{PeNC(Cs_{\mathit{n}}Pb_{\mathit{n}}Cl_{\mathit{m}}Br_{3\mathit{n}-\mathit{m}})^{+}} + e^{-}$$

$$\mathrm{CHCl_3} + e^{-} \longrightarrow \mathrm{CHCl_2}\cdot + \mathrm{Cl^-}$$

$$\mathrm{PeNC(Cs_{\mathit{n}}Pb_{\mathit{n}}Cl_{\mathit{m}}Br_{3\mathit{n}-\mathit{m}})} + \mathrm{Cl^-} \rightleftharpoons \mathrm{PeNC(Cs_{\mathit{n}}Pb_{\mathit{n}}Cl_{\mathit{m}+1}Br_{3\mathit{n}-\mathit{m}-1})} + \mathrm{Br^-}$$

The first two processes are irreversible and may be combined into a single reaction:

$$\mathrm{PeNC(Cs_{\mathit{n}}Pb_{\mathit{n}}Cl_{\mathit{m}}Br_{3\mathit{n}-\mathit{m}})} + \mathrm{CHCl_3} + h\nu \longrightarrow \mathrm{PeNC(Cs_{\mathit{n}}Pb_{\mathit{n}}Cl_{\mathit{m}}Br_{3\mathit{n}-\mathit{m}})^{+}} + \mathrm{CHCl_2}\cdot + \mathrm{Cl^-}$$

The rate of formation of chloride anion in this case would be:

$$\frac{\mathrm{d[Cl^-]_{free}}}{\mathrm{dt}} = \mathrm{k[PeNC][CHCl_3]}$$

The anion species within the PeNC lattice follows a stoichiometric relationship. The exchange of anions between those inside and outside of the PeNC lattice is a reversible process. Consequently, before reaching equilibrium, the reaction rate can be described as follows:

$$-\frac{\mathrm{d[Cl^-]_{free}}}{\mathrm{dt}} = \mathrm{k'[PeNC]_{Br}[Cl^-]_{free}} - \mathrm{k''[PeNC]_{Cl}[Br^-]_{free}}$$

$$\frac{d[Br^-]_{free}}{dt} = k'[PeNC]_{Br}[Cl^-]_{free} - k''[PeNC]_{Cl}[Br^-]_{free}$$

Although $[PeNC]_{Cl}$ and $[PeNC]_{Br}$ are distinguished in this expression, they are treated as the same species in the actual reaction. This is because the excess chloride or bromide anions remaining in the semiconductor nanocrystal can undergo further exchange.

$$\frac{d[Br^-]_{free}}{dt} = -\frac{d[Cl^-]_{free}}{dt} = -k[PeNC][CHCl_3]$$

This rate equation represents the forward reaction of anion substitution, which can be described as a 'bimolecular' reaction involving PeNC and chloroform. The description is consistent with the kinetics dependence on the toluene-to-chloroform ratio and the PeNC concentration, as reported and discussed in the main text.

At the beginning of the experiment, the excess amount of chloroform, acting both as a reactant and a solvent, allows the simplification of the second-order kinetics to pseudo-first-order kinetics. Indeed, a good mono-exponential fit can be obtained for the initial part of the reaction, as shown in Fig. S13. However, as the reaction progresses, several side reactions, including the decomposition of the PeNCs, begin to occur. These side reactions may also be potentially induced by the photoreaction byproducts, as outlined below:

$$CHCl_2Br + e^- \longrightarrow CHCl_2 + Br^-$$

$$R - NH_3{}^+ + Cl^- \longrightarrow R - NH_3Cl$$

As a result, the mono-exponential model deviates as the AI-TA measurement progresses, and the final PB peak position reflects the complex dynamic equilibrium between various side reactions. Note that the charge compensation of ejected electrons is not described here. Although PeNC can be regarded as an electron reservoir, the balance should be completed through an undescribed side reaction that may be involved in this process.

## Supplementary Figures

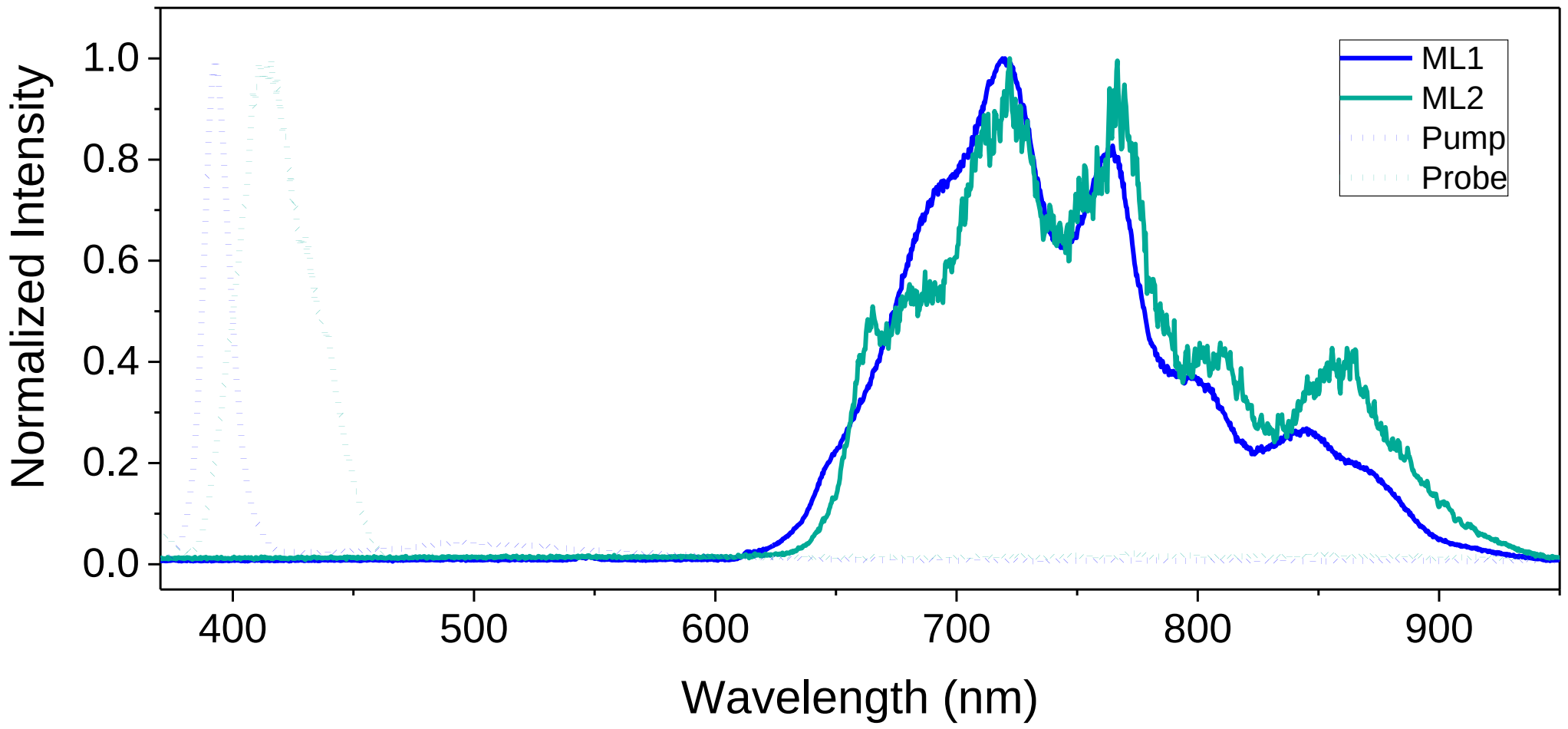

**Fig. S1.** The spectra of ML1 and ML2 oscillators and the spectra of pump and probe pulses generated by the second harmonic generation processes through two different BBO crystals. The data shown is acquired using a grating spectrometer (FLAME-S-VIS-NIR-ES, Ocean Optics). The spectra of fundamental outputs are also shown in the near-infrared region. The critical phase-matching condition of BBO crystal is adjusted to generate different pump and probe spectra in the near-ultraviolet (NUV) region.

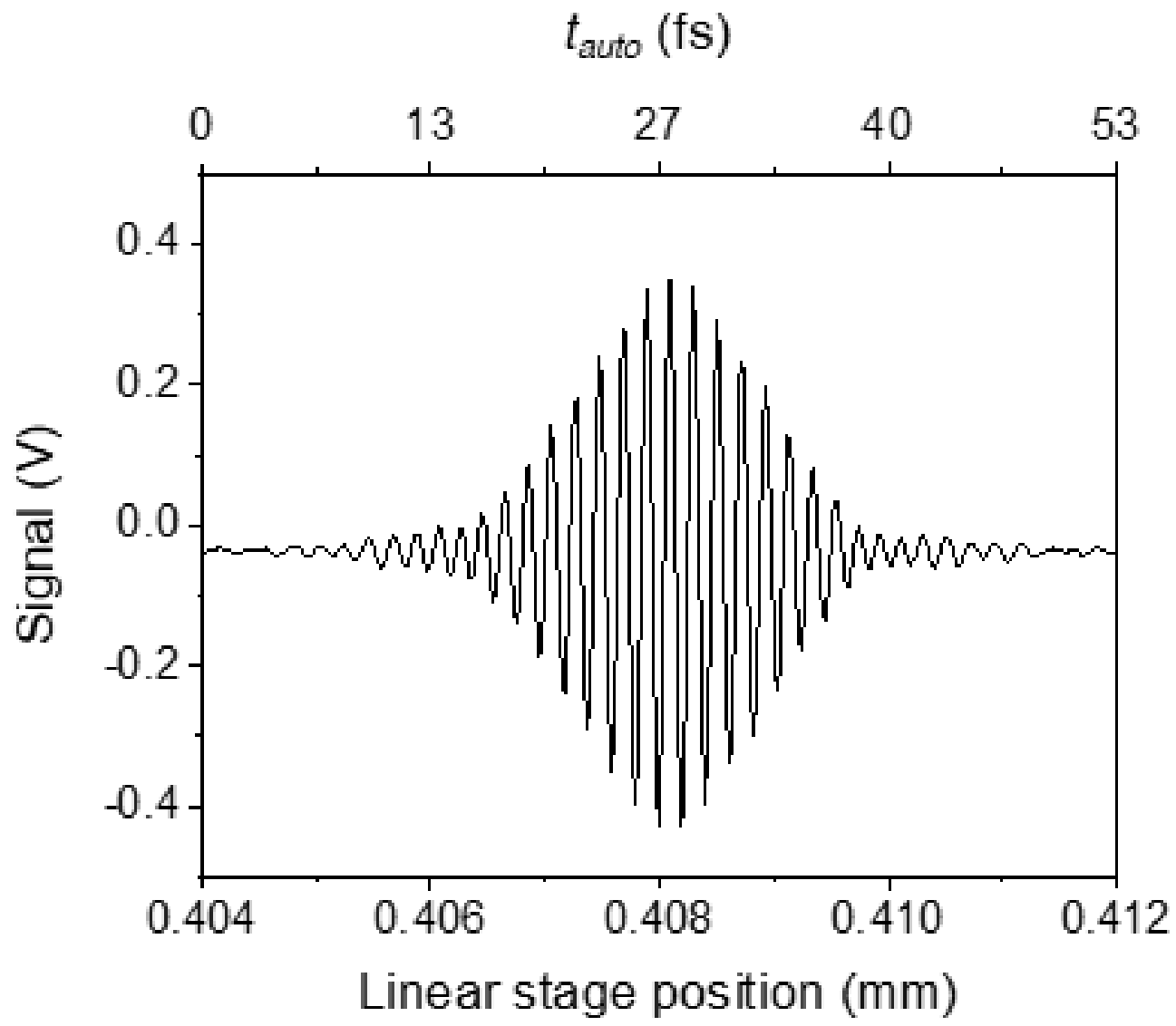

**Fig. S2.** The interferogram generated from the scanning of the linear delay stage. The position in the stage can be converted into pulse delay between split probe pulses by dividing it with the speed of light. The FWHM of the autocorrelation signal is approximately 13.5 fs. Here, the linear stage is employed to operate in the range of pulse duration. The short scan length of 8 μm distinguishes its purpose from the linear delay stage used in the conventional pump-probe technique, which is utilized to generate delay between pump and probe pulses and may scan up to tens of centimeters.

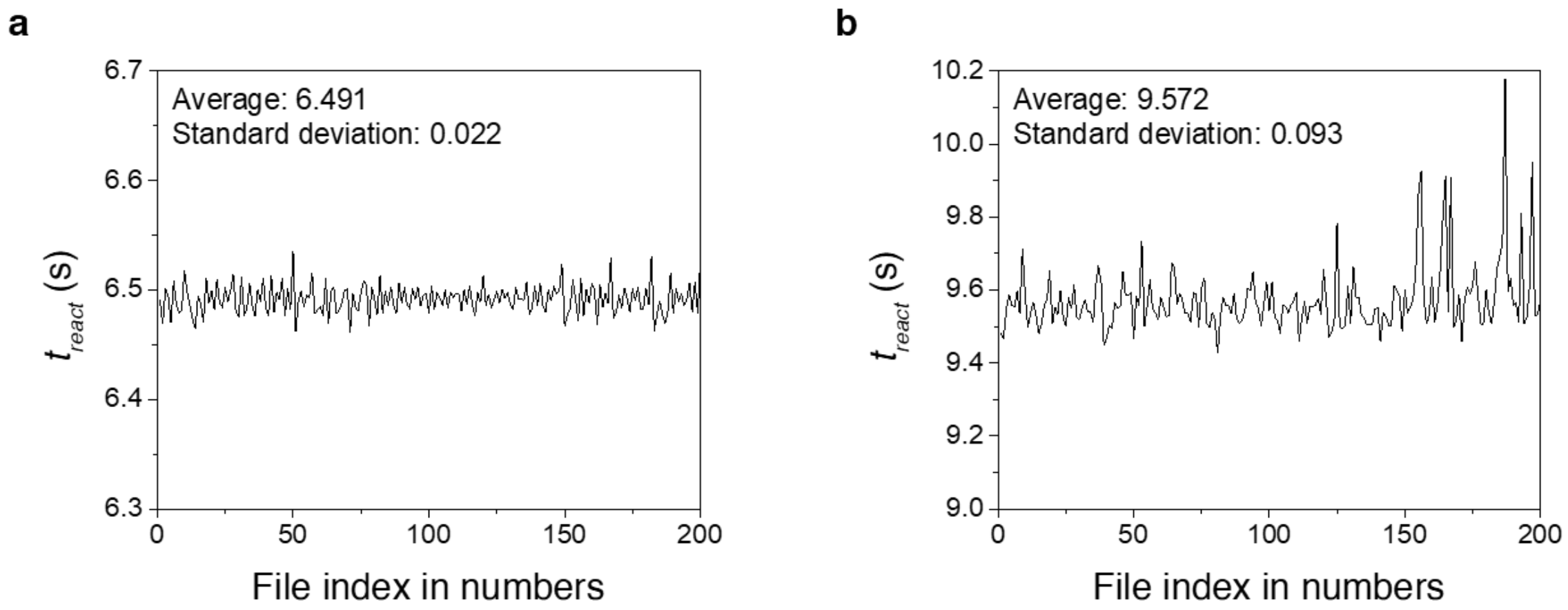

**Fig. S3.** The time interval between recorded datasets for acquisition (**a**) up to 15 ps and (**b**) up to 1 ns, which is calculated by subtracting the recorded time stamps of consecutive saved files. Note that the data acquisition time gets longer due to the more extensive computational resources required for processing the data and not due to actual experimental procedures (i.e., interferogram scanning). Windows OS provides timestamps with an accuracy of up to milliseconds, and the reliability of time stamps can be improved by installing time reference programs allowing synchronization with Coordinated Universal Time (UTC) provided by KRISS (Korea Research Institute of Standard and Science).

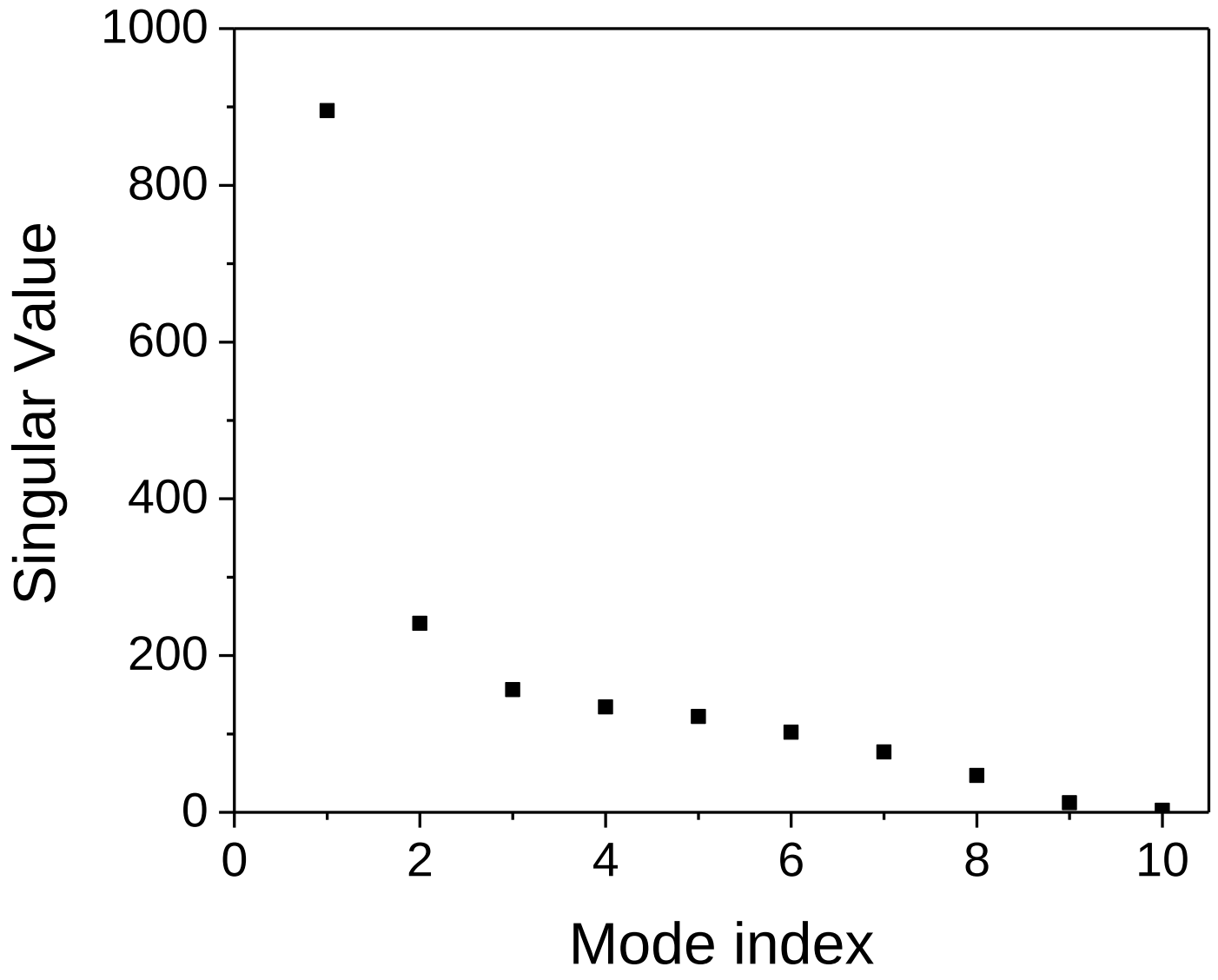

**Fig. S4.** Singular values arranged from high to low acquired from the analysis of the transient absorption spectrum in Fig. 2.

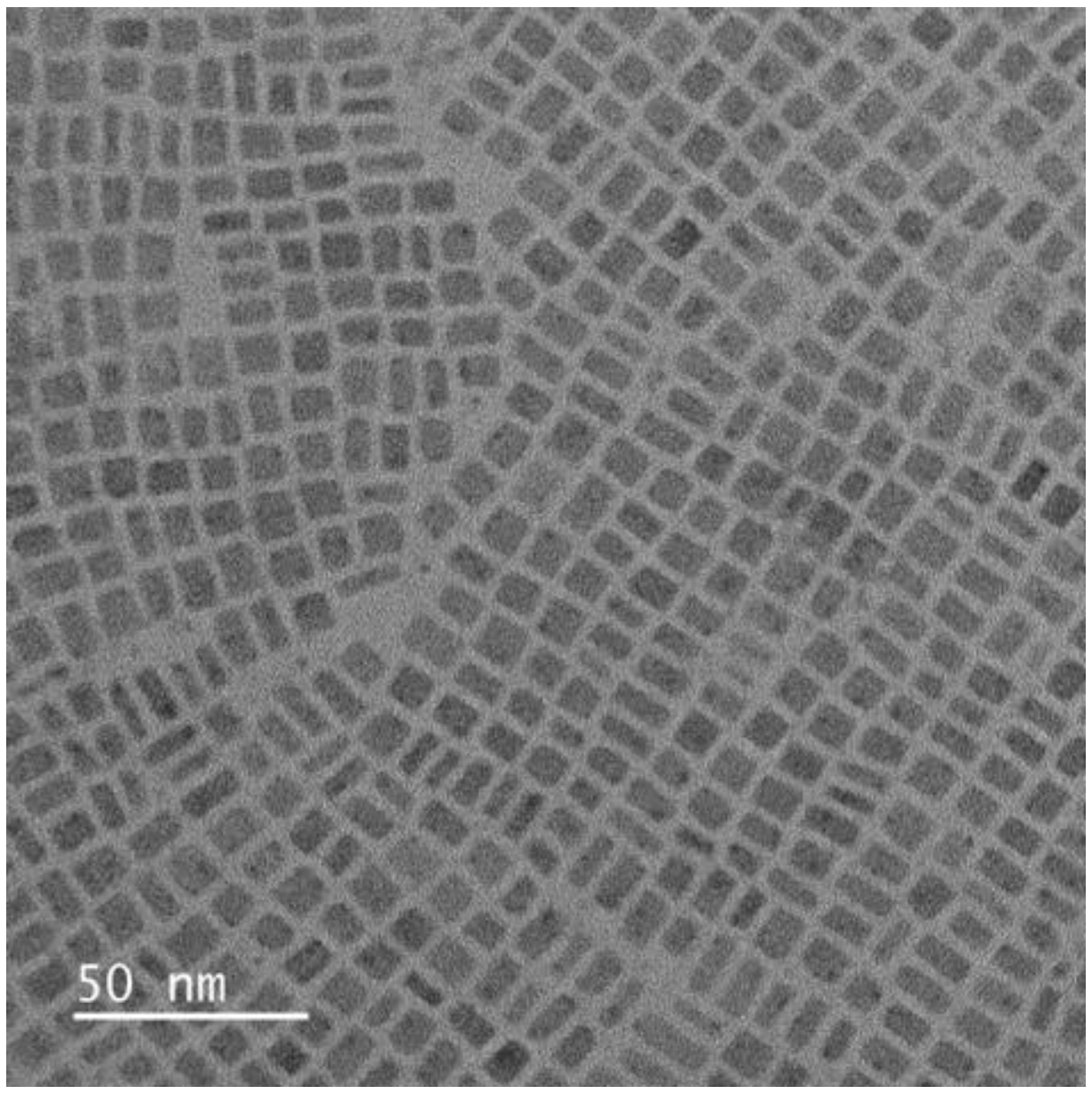

**Fig. S5.** TEM image of the $CsPb(Cl,Br)_3$ PeNCs. The scale bar is 50 nm. On average, the particle size is 7.2 nm, well below 10 nm in size, indicating a strong confinement regime[42].

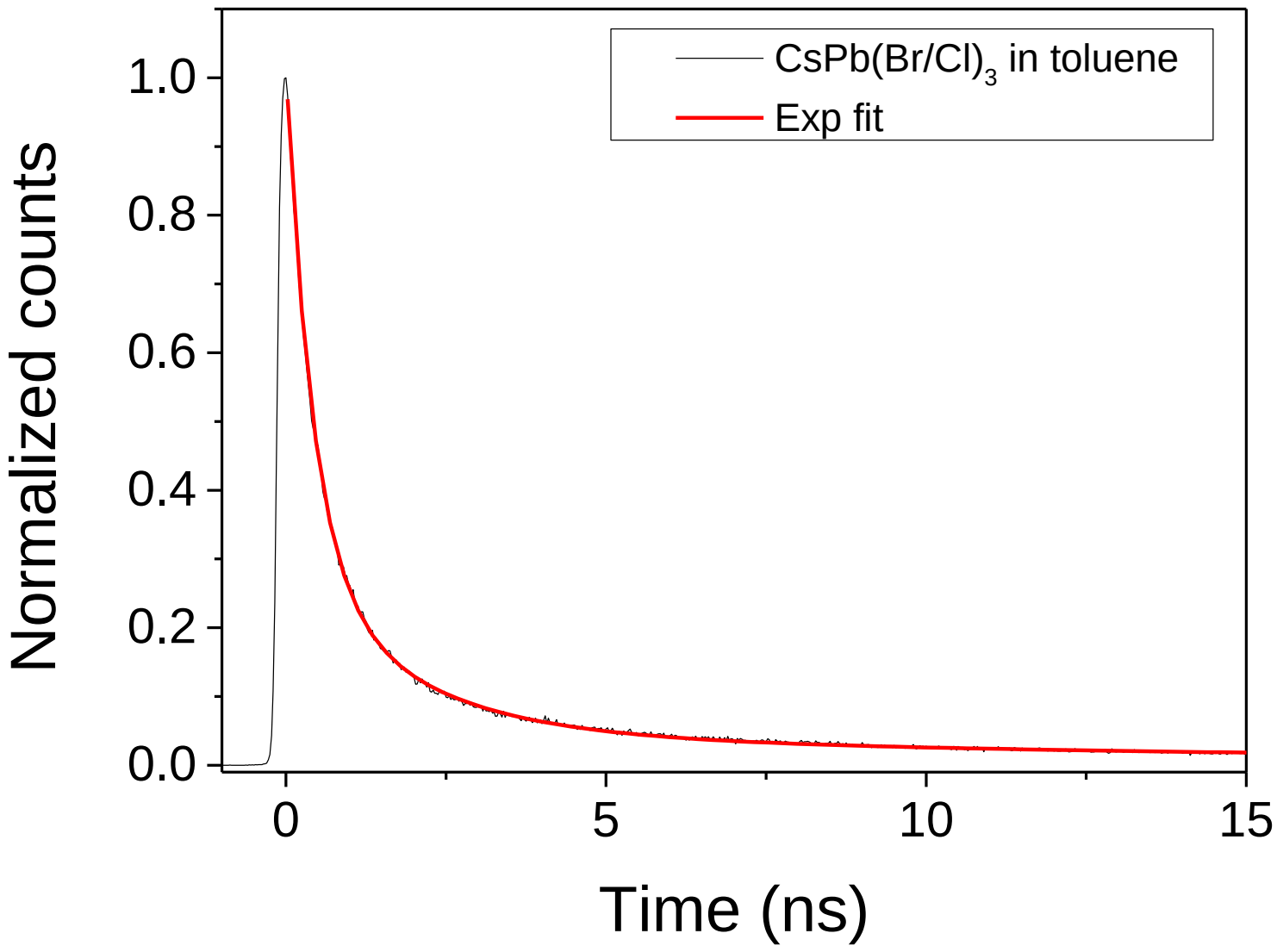

**Fig. S6.** Time-correlated single photon counting result of PeNC solution dissolved in toluene. The time bin is set to 25 ps. The sample is excited with 398 nm light with a 1.7 nm bandpass filter, and emitted photons are detected at 440 nm with a 5 nm bandpass filter. The fitting presented here results from a tri-exponential fit, and the weighted average lifetime is 424.7 ps.

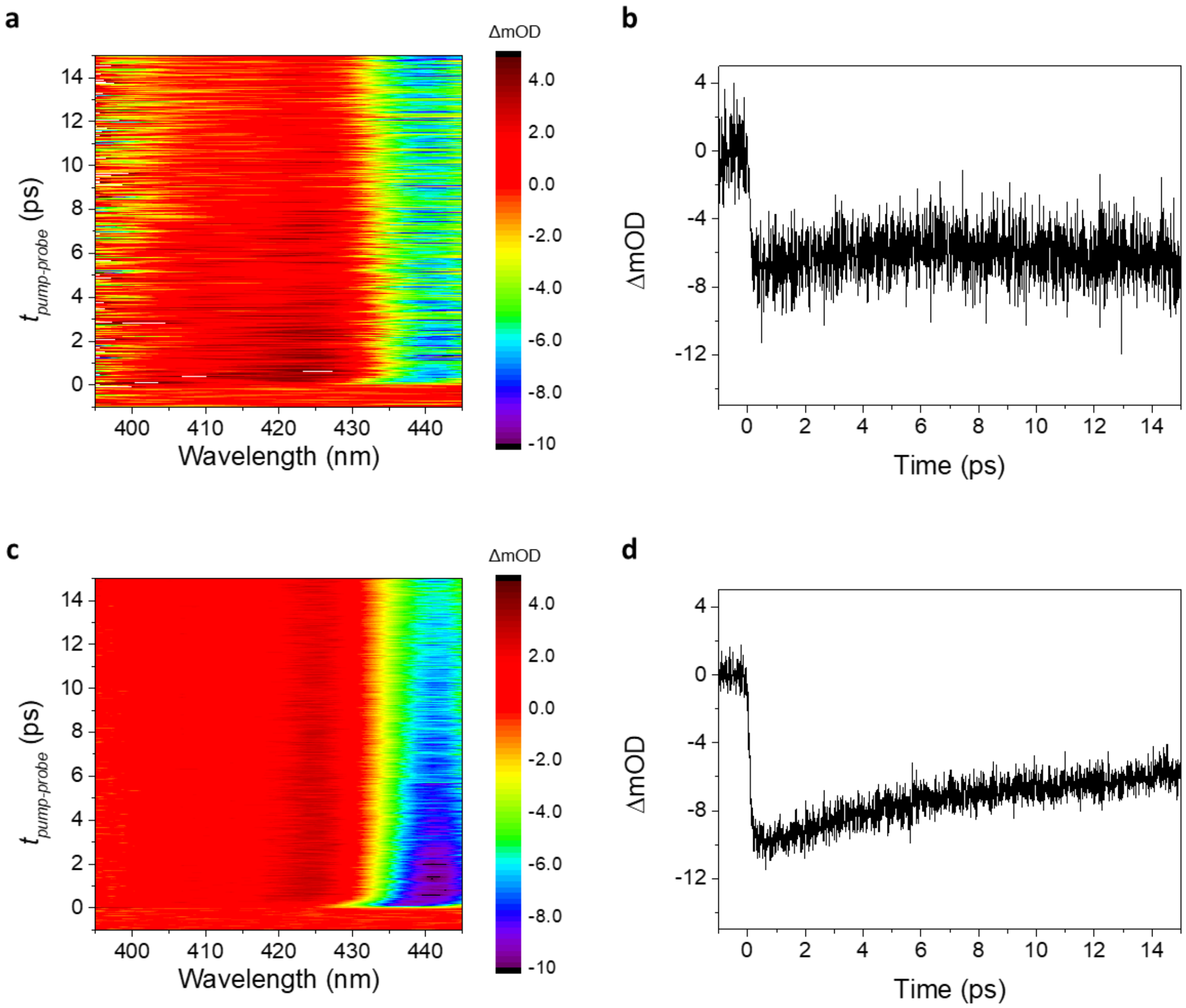

**Fig. S7.** (a) An exemplary TA result of a single scan interferogram and (b) its time profile at 440.8 nm. Noise is prominent, and only strong transient features can be identified. (c) When FFT of 10 interferometric scans are averaged, the spectral features and (d) the lifetime profile are better defined. The scans with too large deviations in time profiles were baseline-corrected before being averaged.

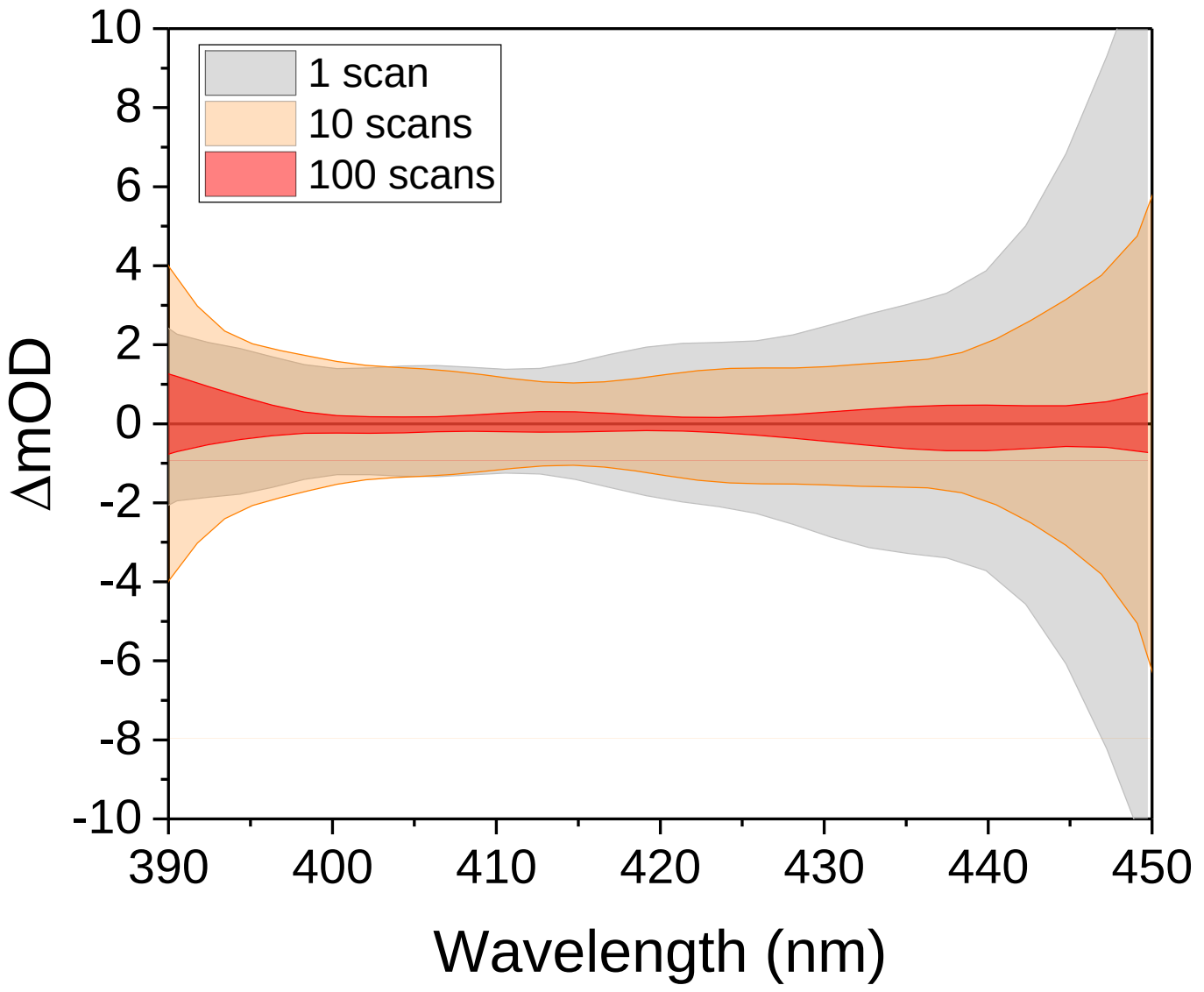

**Fig. S8.** Standard deviations depending on the averaging numbers as a function of wavelength.

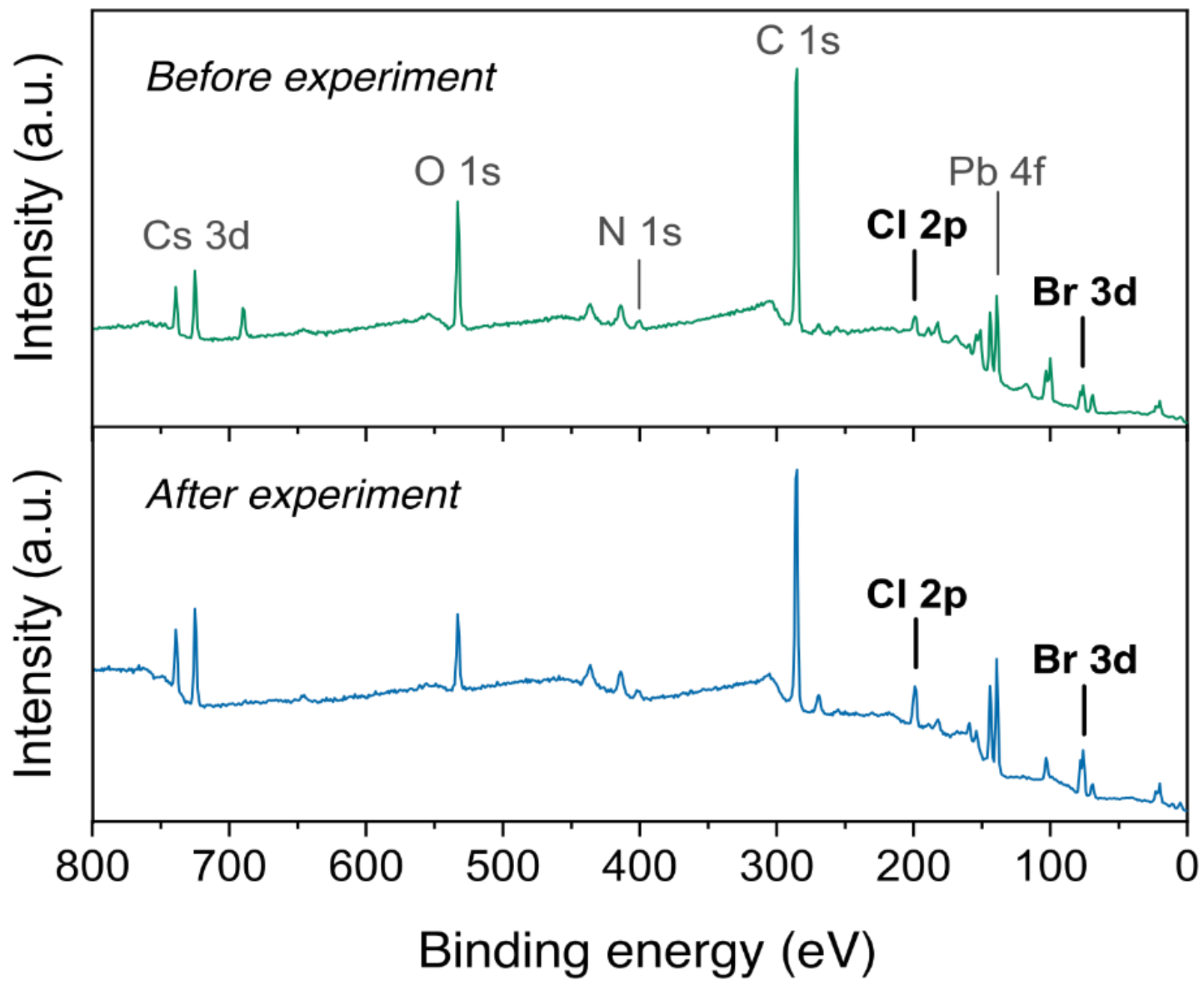

**Fig. S9.** XPS wide scan spectra of $CsPb(Br/Cl)_3$ PeNCs before and after photoinduced anion exchange reaction. The unidentified peaks are stray peaks that do not belong to $CsPb(Br/Cl)_3$ PeNCs.

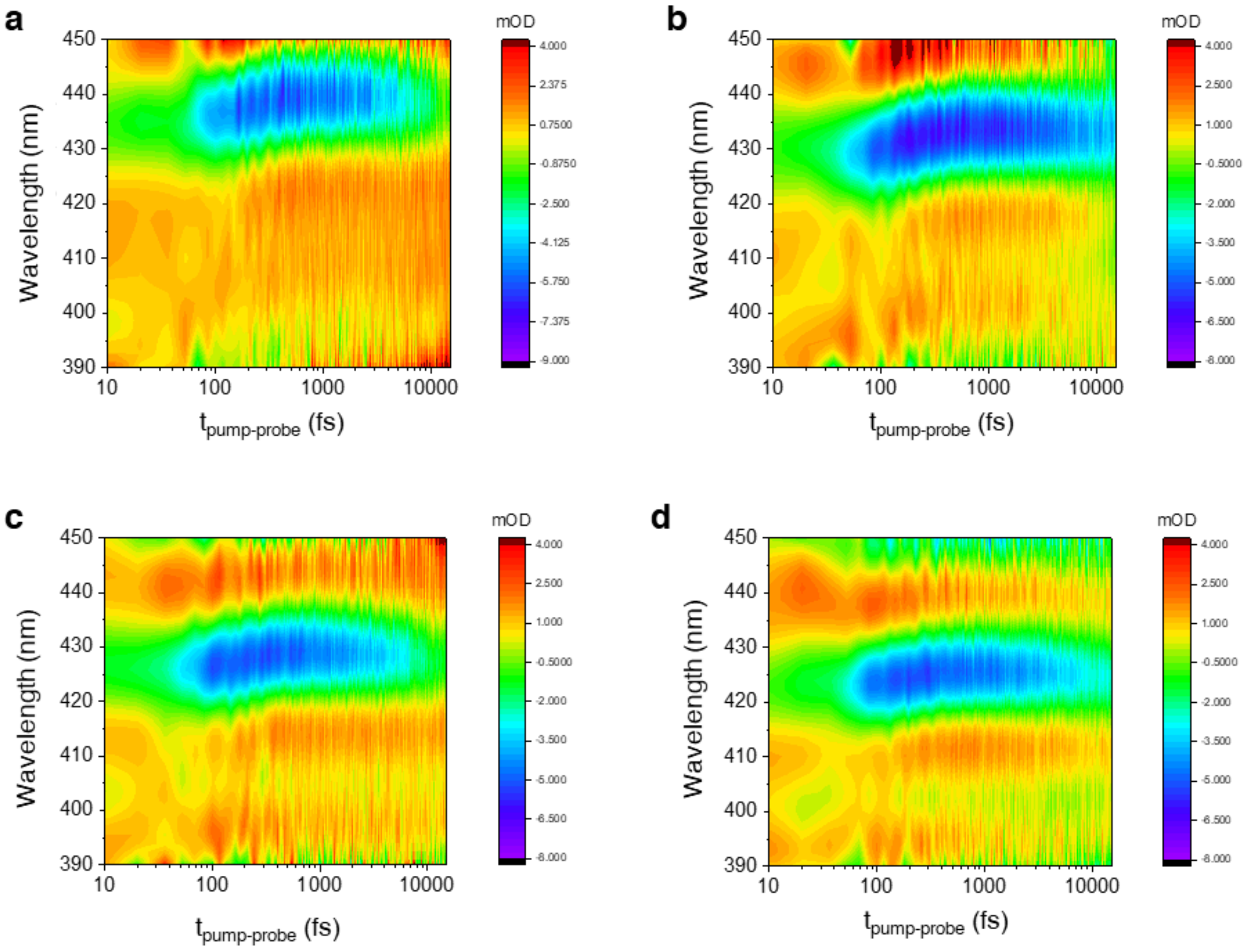

**Fig. S10.** Selected transient absorption of photo-substitution AI-TA measurement of $CsPb(Cl/Br)_3$ nanoparticles at (**a**) the beginning of the experiment (0 – 1 min), (**b**) after 10 minutes, (**c**) after 20 minutes and (**d**) after 29 minutes.

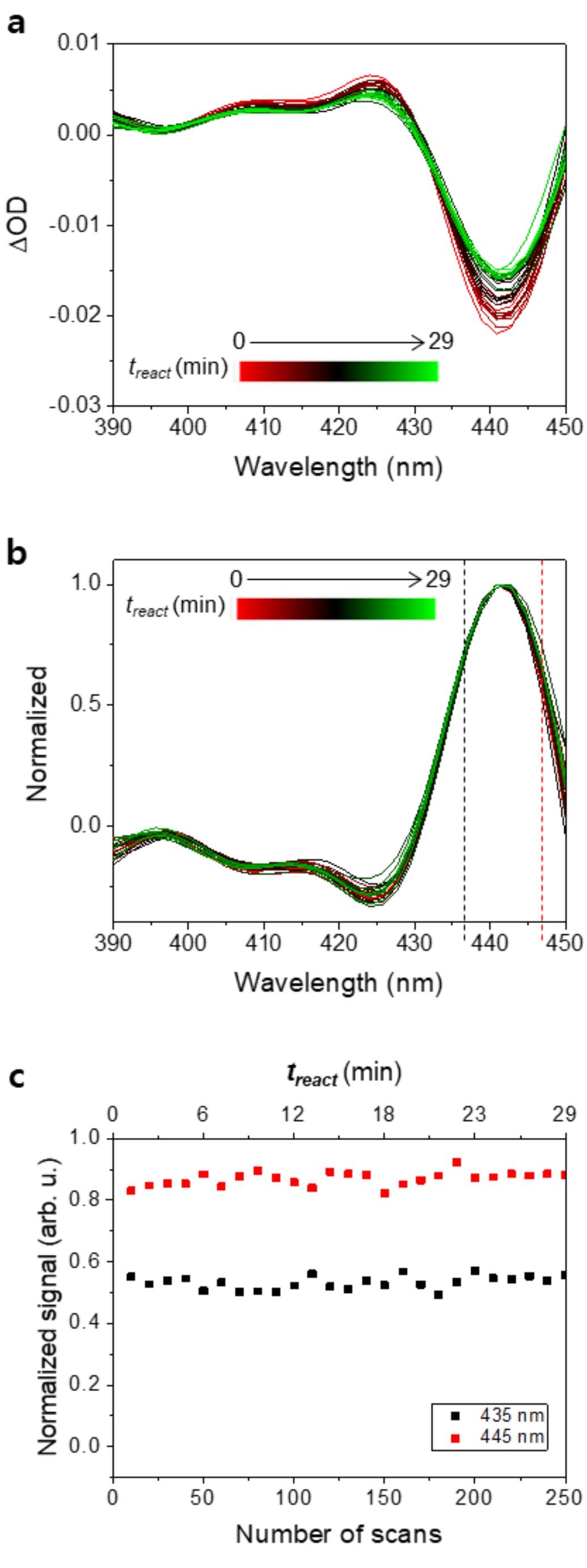

**Fig. S11. (a)** Temporal variations of TA spectra at 1 ps of PeNC dissolved in toluene for 30 minutes. **(b)** Normalized TA spectra of (a) and **(c)** the amplitude tracking of normalized values at a fixed wavelength of 435 nm and 445 nm. The signal loss is attributed to the sample decomposition.

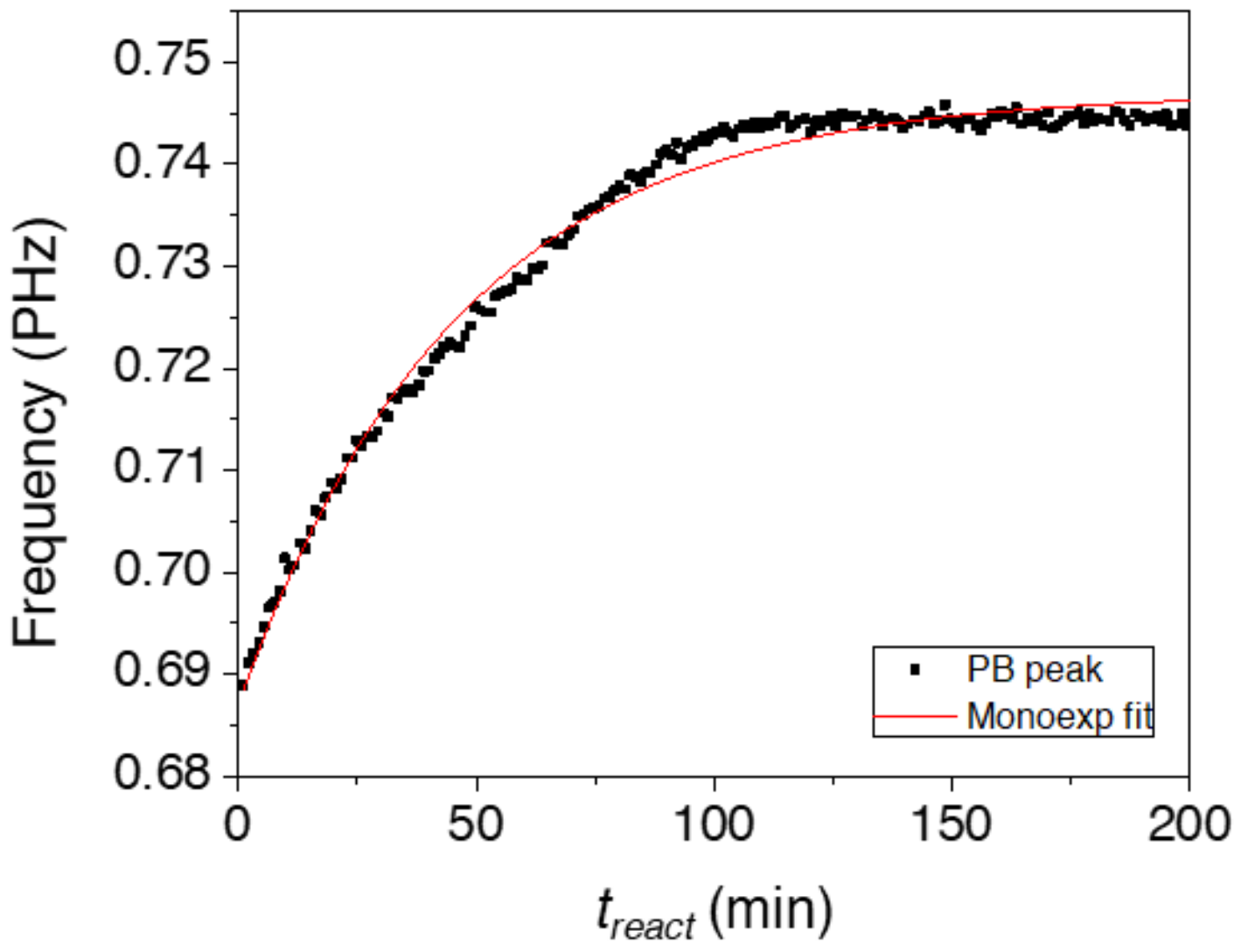

**Fig. S12.** Kinetic analysis of transient absorption peak shift by fitting it with mono-exponential function. The wavelength axis originally presented in Figure 4a is converted to the frequency axis here to ensure an even energy spacing. The lifetime is determined to be 46.0 min, with an R-square of 0.989. The deviation of the fit as it progresses is discussed in Supplementary Note 4.

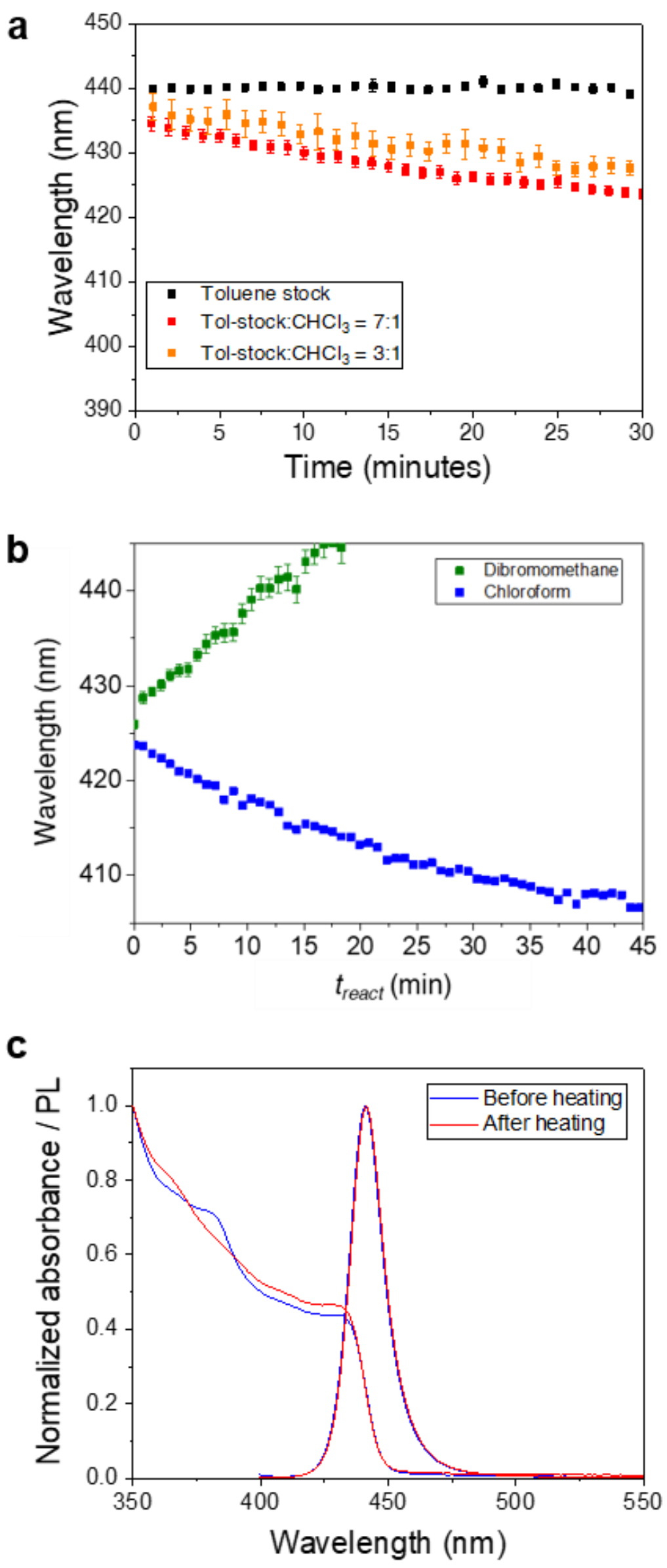

**Fig. S13.** Effect of halide species and light on the AI-TA experiment. (**a**) Effect of halide solvent on PB peak, by comparing peak positions in toluene-only and a toluene/chloroform mixed solvent. It is shown that the smaller ratio of PeNC stock solution (Tol-stock) is shown to slow down the rate of substitution, in spite of the increased chloroform ratio. (**b**) Starting with PeNC, which possesses a bandgap near 425 nm through the control of the halide ratio during the synthesis step, AI-TA is conducted after adding chloroform / dibromomethane to the mixture. Two separate experiments are plotted on one graph as a function of reaction time ($t_{react}$). (**c**) Comparison of absorption and emission spectra of PeNC in chloroform before and after heating the solution at 100 °C for two hours in the dark.

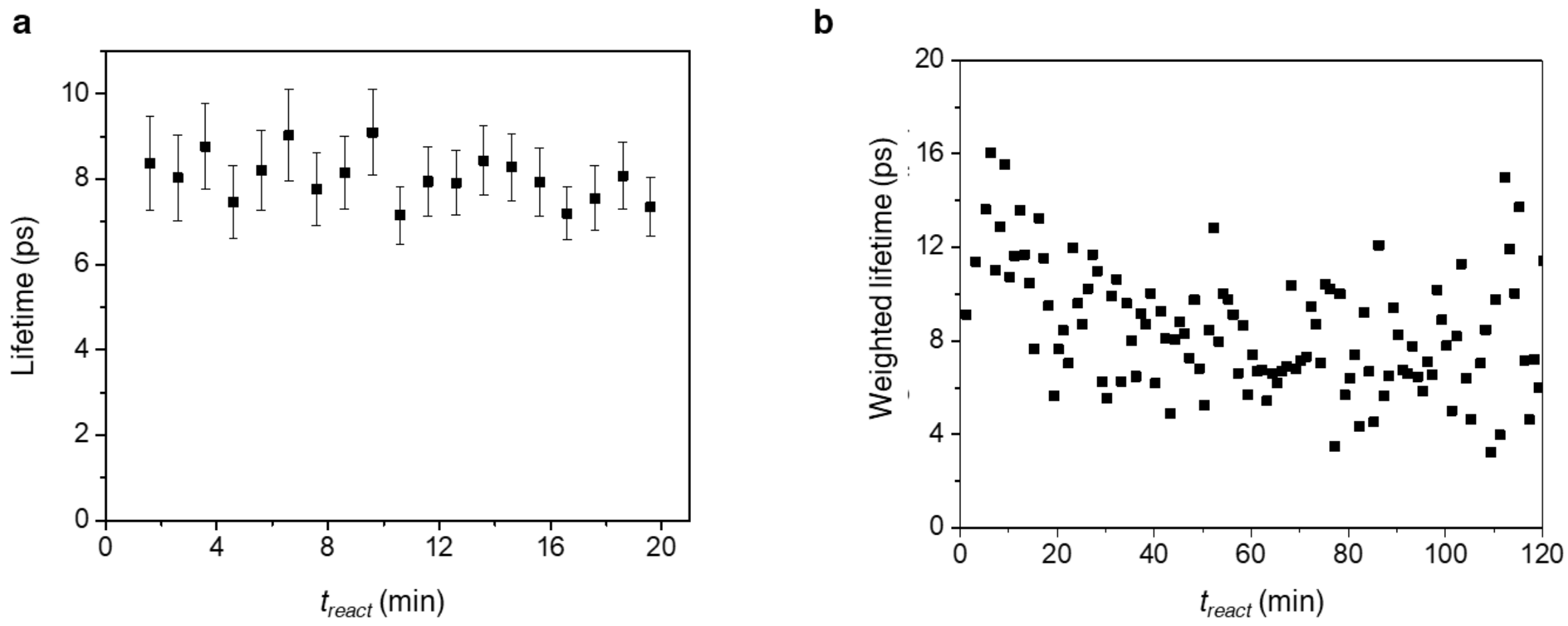

**Fig. S14.** Early decay component (fit from < 50 ps) of PeNCs as AI-TA measurement progresses. (**a**) Fit result of experiment in Fig. 3b and (**b**) fit result of experiment presented in Fig. 4a.

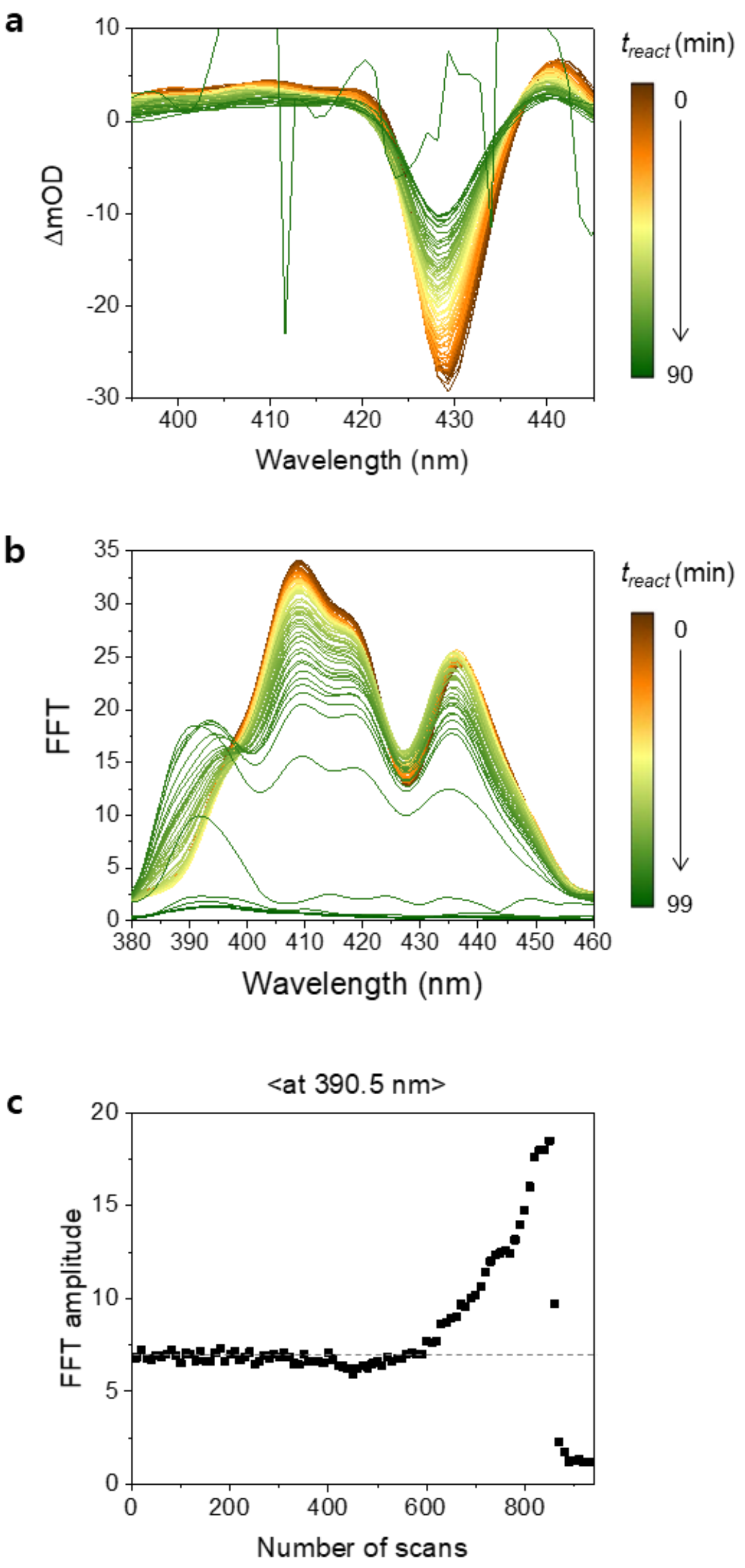

**Fig. S15.** (a) Full collection of TA spectra at 1 ps for 2ML NPL experiment done over 90 minutes (the truncated spectra up to 42.3 minutes is presented in Fig. 5e). The last scan around 90 minutes is reduced to noise due to aggregation at the focus. (b) FFT spectra of AI-TA data up to 99 minutes of reaction time, at negative time delays (before pump is incident). The figure shows increasing values near 427 nm, which represents a reduction of sample absorption due to degradation. (c) After approximately 400 scans, the pump scattering centered at 390 nm begins to appear.

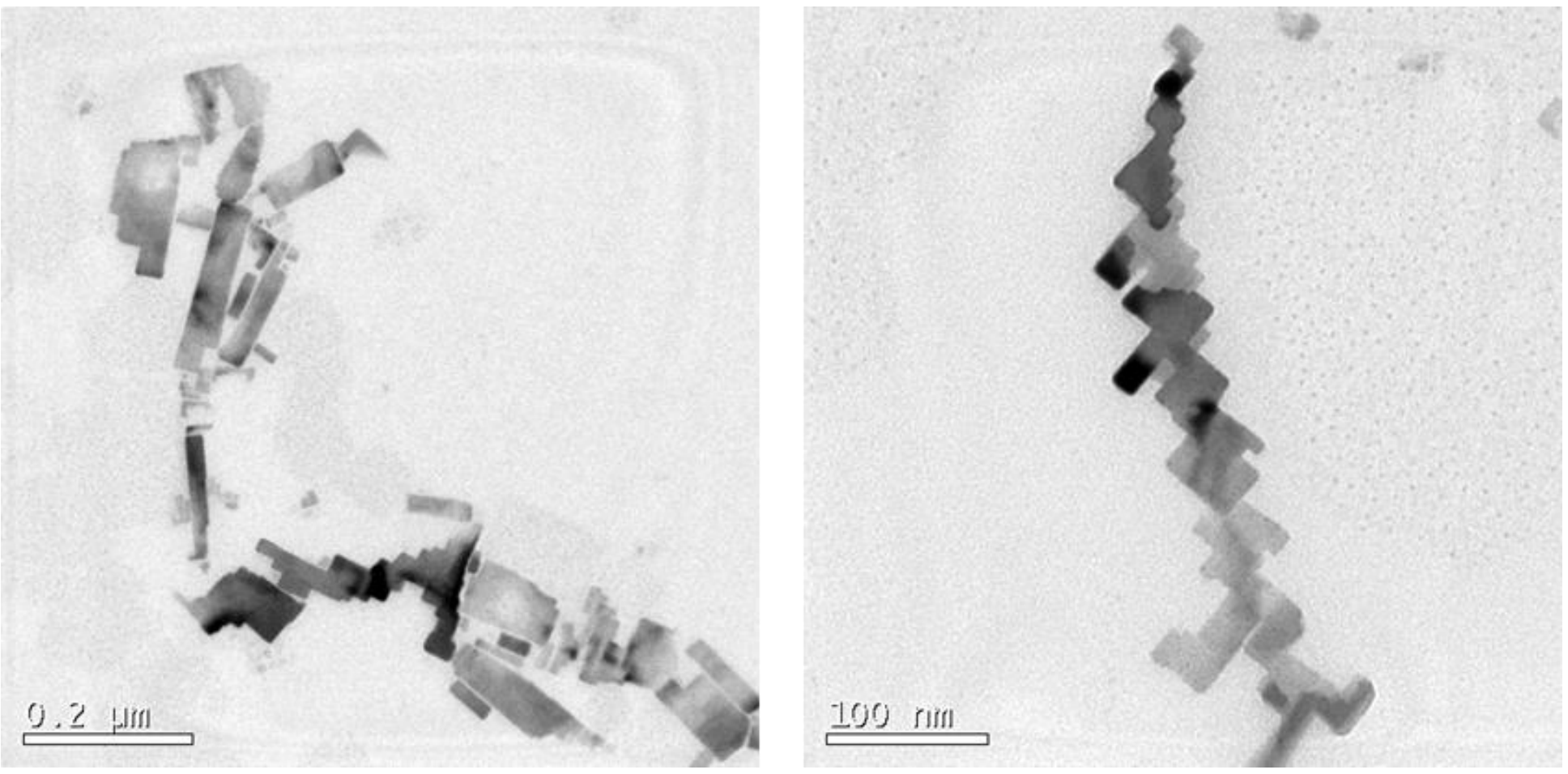

**Fig. S16.** Image of mosaic-like structure found in TEM images of PeNPL sample after the experiment.

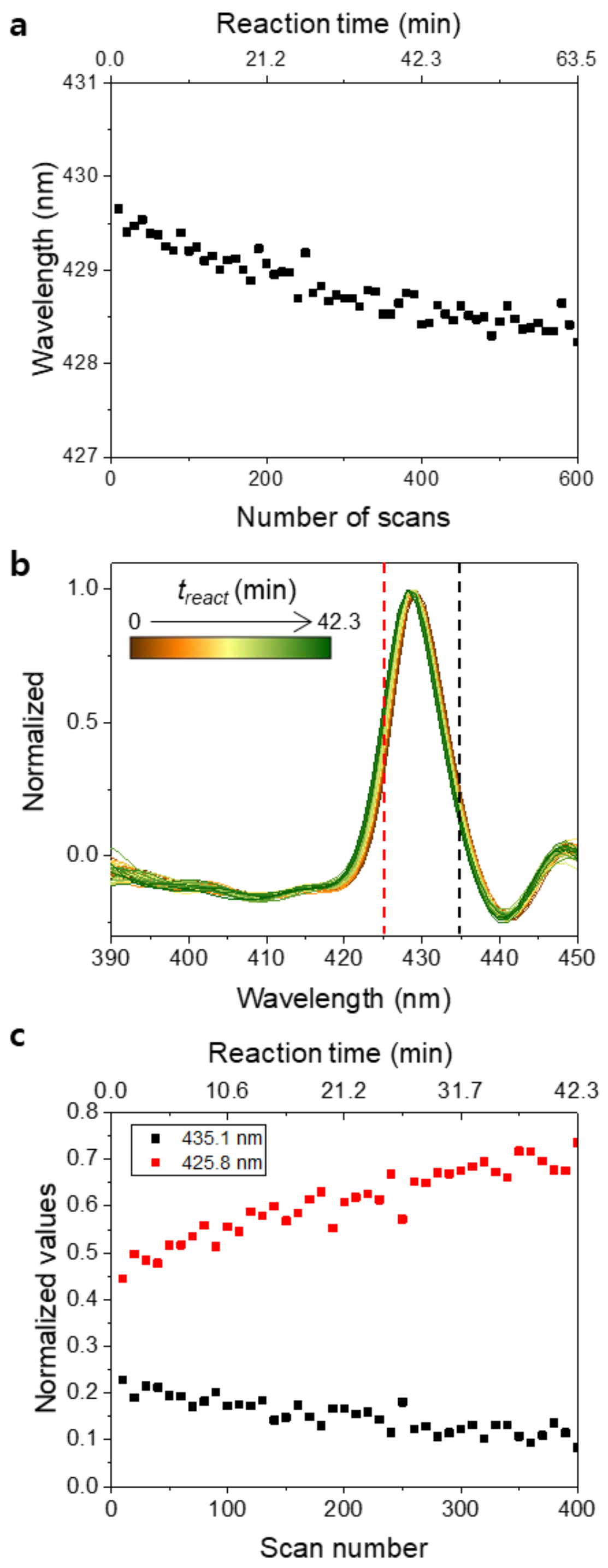

**Fig. S17. (a)** Values of PB peak acquired by Gaussian fitting of the spectra shown in Figure 4e. **(b)** Collections of normalized spectra at 1 ps until 400 scans. (**c**) The amplitude tracking of normalized spectra values at a fixed wavelength of 435 nm and 425 nm.

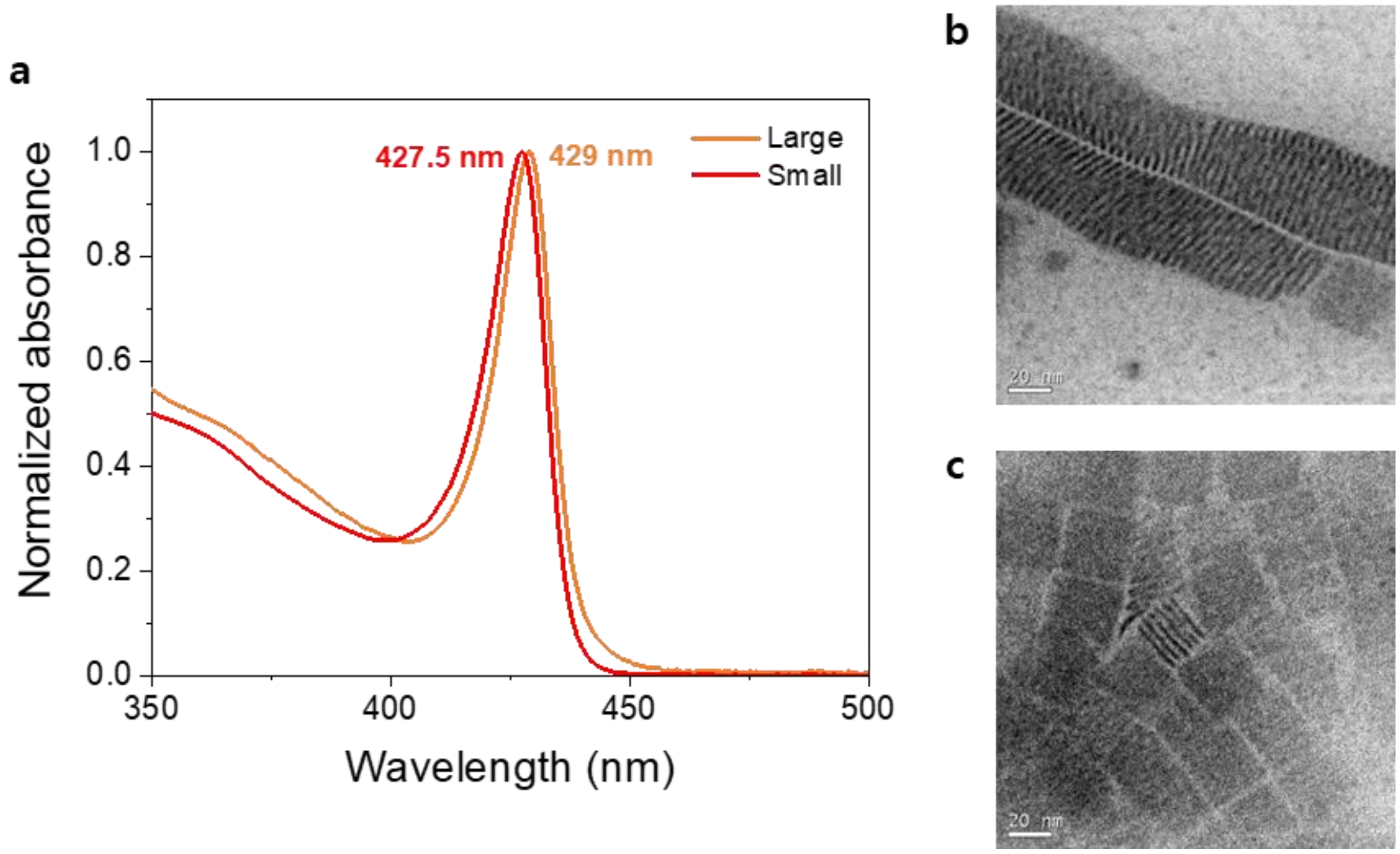

**Fig. S18. (a)** Absorption spectra comparing PeNPLs larger and smaller in nonconfined dimension. (**b**) TEM images of large PeNPLs and (**c**) small PeNPLs. The scale bar is 20 nm.

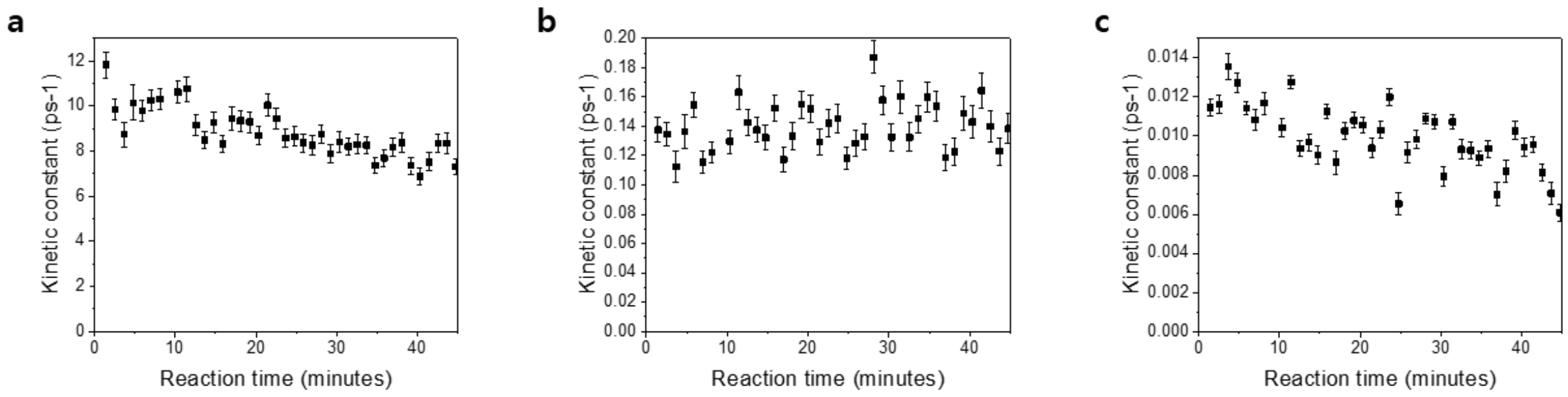

**Fig. S19.** Three lifetimes acquired using exponential fit, representing (a) subpicosecond hot carrier cooling, (b) charge trapping process, and (c) charged exciton recombination. Data points contaminated by significant noise were omitted from the fitting analysis.

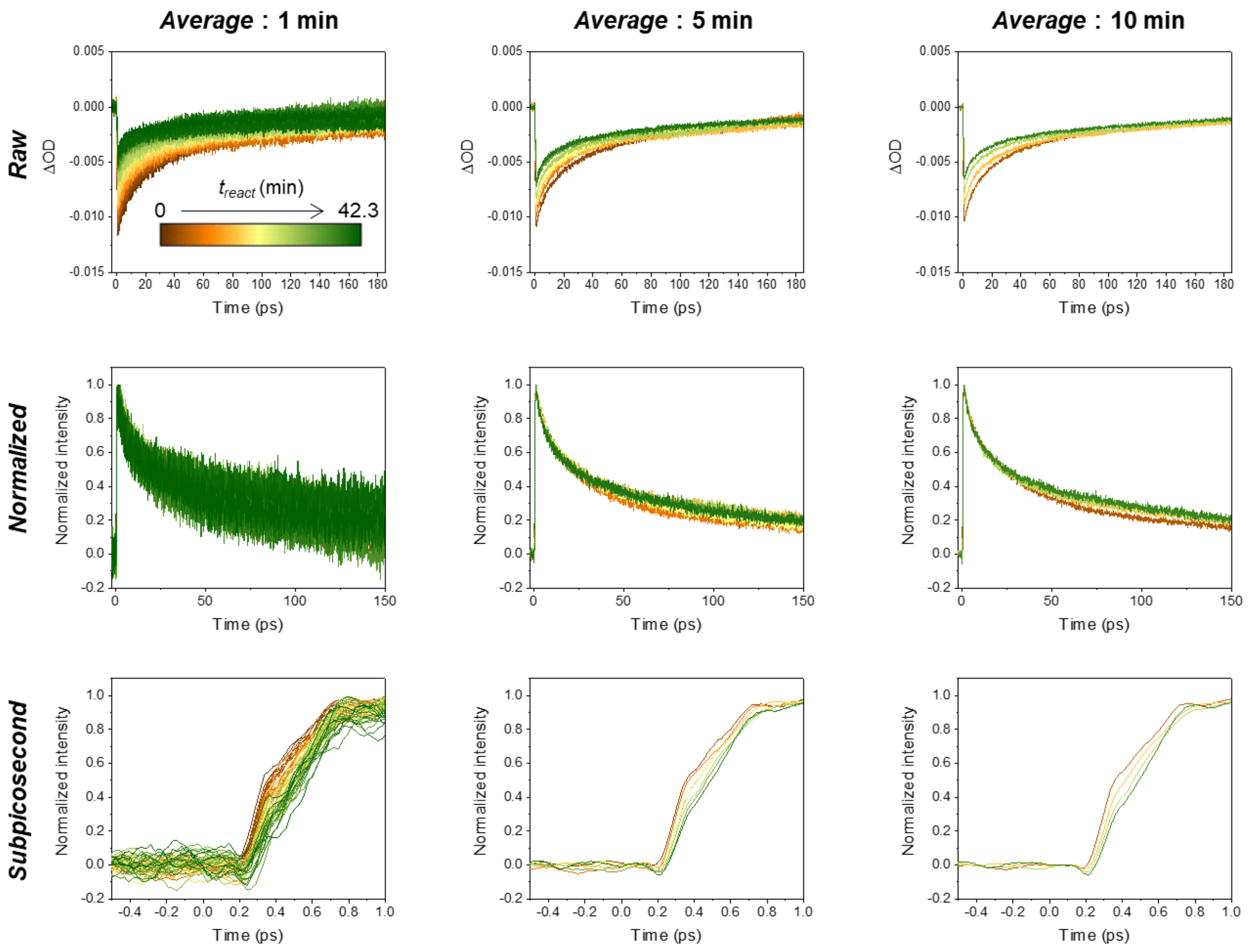

**Fig. S20.** Collections of time profiles at 431.6 nm of data shown in Figure 4d. Different lengths of averaging $t_{react}$ segments allow clearer visualization of changes in temporal profiles. As the measurement progresses, a delayed rise, unchanging kinetics up to 30 ps, and a delayed decay afterward are observed.

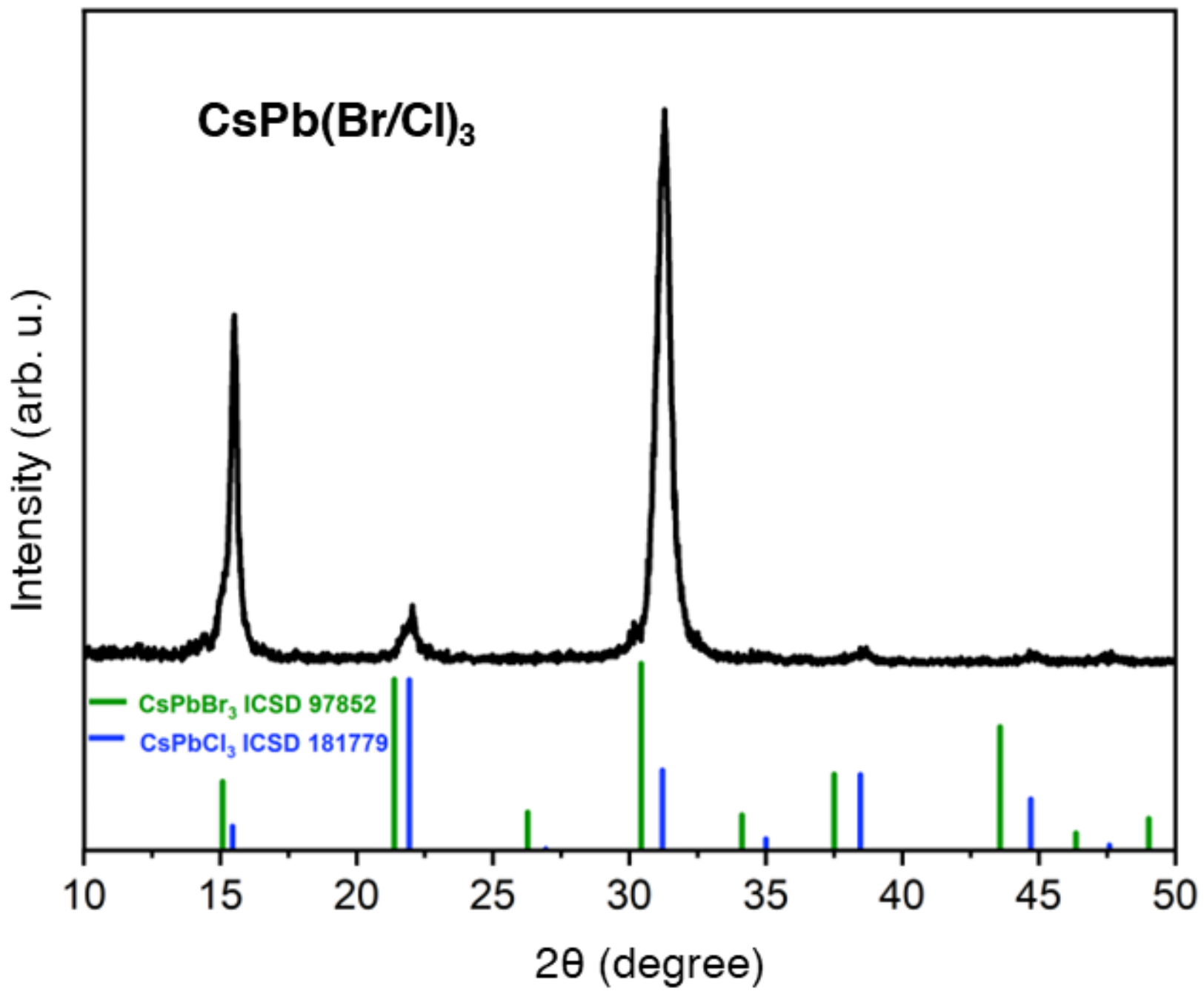

**Fig. S21.** XRD spectra of $CsPb(Br/Cl)_3$ PeNCs and references bulk $CsPbBr_3$ (ICSD 97852) and $CsPbCl_3$ (ICSD 181779). The diffraction peaks of $CsPb(Br/Cl)_3$ exhibited shifting to a high angle (2theta) compared to the reference bulk peaks, indicating the formation of mixed halide PeNCs.

**Table S1.** Reduced representation of high-order singular value decomposition (HOSVD) result. The three variables shown in $M(P_1, P_2, P_3)$, indicate effective time, wavelength, and experimental time, respectively.

| $P_3$: 1 | | $P_1$ | | | | | |
|---|---|---|---|---|---|---|---|
| | | 1 | 2 | 3 | 4 | 5 | 6 |
| $P_2$ | 1 | 5032.8 | 42.2 | 46.6 | 2.9 | 0.2 | 3.9 |
| | 2 | 15.9 | 197.3 | 120.2 | 3.4 | 2.9 | 0.6 |
| | 3 | 4.1 | 82.7 | 12.1 | 11.5 | 9.7 | 1.7 |
| | 4 | 16 | 53.1 | 58.1 | 1.9 | 6.4 | 18.2 |
| | 5 | 1.9 | 78.2 | 45.3 | 7.2 | 9.9 | 8.7 |
| | 6 | 26.3 | 52.9 | 58.2 | 56.8 | 21 | 10 |

| $P_3$: 2 | | $P_1$ | | | | | |
|---|---|---|---|---|---|---|---|
| | | 1 | 2 | 3 | 4 | 5 | 6 |
| $P_2$ | 1 | 21.4 | 41.7 | 139.3 | 2.6 | 5.1 | 2 |
| | 2 | 3593.3 | 8.2 | 6.1 | 2.5 | 2.4 | 11.3 |
| | 3 | 69.1 | 157.9 | 87.2 | 2.7 | 0.5 | 10.4 |
| | 4 | 49.1 | 11.1 | 11.1 | 5.6 | 6.3 | 8 |
| | 5 | 3.2 | 57.4 | 30.7 | 2.2 | 4.9 | 8.8 |
| | 6 | 0.4 | 87.8 | 25.9 | 1.4 | 8.1 | 27.7 |

| $P_3$: 3 | | $P_1$ | | | | | |
|---|---|---|---|---|---|---|---|
| | | 1 | 2 | 3 | 4 | 5 | 6 |
| $P_2$ | 1 | 7.2 | 67.9 | 66.1 | 0.9 | 5 | 4.1 |
| | 2 | 77.1 | 23.3 | 121.7 | 8.9 | 10.7 | 4.9 |
| | 3 | 3151.6 | 54 | 17.9 | 0.6 | 7.5 | 12.5 |
| | 4 | 14.5 | 89.7 | 98.4 | 0.8 | 2.3 | 14.5 |
| | 5 | 31.6 | 143.1 | 4.4 | 3.7 | 7.1 | 6.2 |
| | 6 | 51 | 99.4 | 56.4 | 8 | 16.9 | 10.7 |

| **$P_3$**: 4 | | **$P_1$** | | | | | |
|---|---|---|---|---|---|---|---|
| | | 1 | 2 | 3 | 4 | 5 | 6 |
| **$P_2$** | 1 | 12.2 | 13.7 | 48.9 | 0.2 | 1.1 | 15.2 |
| | 2 | 38.4 | 28.8 | 25.2 | 2.4 | 3.8 | 13.5 |
| | 3 | 25 | 8.6 | 79 | 4.3 | 6.1 | 7.9 |
| | 4 | 2485 | 16.2 | 61 | 2.1 | 7.4 | 12.7 |
| | 5 | 106.5 | 77.7 | 27.2 | 2.4 | 4.1 | 0.6 |
| | 6 | 24.1 | 61.2 | 15.8 | 0.8 | 16.5 | 25.3 |

| **$P_3$**: 5 | | **$P_1$** | | | | | |
|---|---|---|---|---|---|---|---|
| | | 1 | 2 | 3 | 4 | 5 | 6 |
| **$P_2$** | 1 | 3.3 | 24.4 | 18.5 | 0.9 | 0.8 | 4.3 |
| | 2 | 5.1 | 38.2 | 7.2 | 0.6 | 1.1 | 1.7 |
| | 3 | 19 | 9.3 | 23.1 | 2.8 | 2.7 | 0.1 |
| | 4 | 45.1 | 13.9 | 6.9 | 2.4 | 4.3 | 15.8 |
| | 5 | 1134.2 | 23.4 | 21.4 | 0.2 | 2.7 | 15.7 |
| | 6 | 93 | 27.2 | 10 | 5 | 5.2 | 1.4 |

| **$P_3$**: 6 | | **$P_1$** | | | | | |
|---|---|---|---|---|---|---|---|
| | | 1 | 2 | 3 | 4 | 5 | 6 |
| **$P_2$** | 1 | 0.421 | 4.15 | 5.31 | 0.0742 | 1.81 | 1.78 |
| | 2 | 0.696 | 2.38 | 0.896 | 0.0381 | 0.793 | 1.23 |
| | 3 | 0.842 | 1.20 | 2.04 | 1.74 | 4.55 | 1.68 |
| | 4 | 4.77 | 5.83 | 8.23 | 1.21 | 0.342 | 2.60 |
| | 5 | 47.7 | 5.44 | 1.44 | 5.01 | 1.31 | 2.19 |
| | 6 | 152 | 1.60 | 8.90 | 2.91 | 25.7 | 30.1 |